\documentclass[aps,prd,nofootinbib,notitlepage,superscriptaddress,11pt]{revtex4-2}

\usepackage{stmaryrd}
\usepackage{comment}
\usepackage{color}
\usepackage{amsmath,bm}
\usepackage{amssymb}
\usepackage{amsthm}
\usepackage{mathrsfs}
\usepackage{graphicx}
\usepackage{marvosym}
\usepackage{array}
\usepackage{latexsym}
\usepackage{enumerate}
\usepackage{slashed}
\usepackage{hyperref}
\usepackage[normalem]{ulem}

\allowdisplaybreaks[1]

\begin{document}

\title{Spin polarization and correlation of quarks from glasma}
\author{Avdhesh Kumar}
\affiliation{
	Institute of Physics, Academia Sinica, Taipei 11529, Taiwan\\
}
\author{Berndt M\"uller}
\affiliation{
	Department of Physics, Duke University, Durham, North Carolina 27708-0305, USA\\
}
\author{Di-Lun Yang}
\affiliation{
	Institute of Physics, Academia Sinica, Taipei 11529, Taiwan\\
}
\date{\today}

\begin{abstract}
We investigate the interaction of strong color fields in the glasma stage of high-energy nuclear collisions with the spins of quarks and antiquarks. We employ the perturbative solution of the quantum kinetic theory for the spin transport of (massive) quarks in a background color field governed by the linearized Yang-Mills equation and derive expressions for the quark-spin polarization and quark-antiquark spin correlation at small momentum in terms of field correlators. For the Golec-Biernat W\"usthoff dipole distribution the quark-spin polarization vanishes, but the out-of-plane spin correlation of quarks and antiquarks is nonzero. Our order-of-magnitude estimate of the correlation far exceeds that caused by vorticity effects, but does not fully explain the data for vector meson alignment. We identify possible mechanisms that could further increase the predicted spin correlation. 
\end{abstract}
\maketitle

\section{Introduction}
Recent observations of global and local spin polarization of hadrons in relativistic heavy ion collisions \cite{STAR:2017ckg,STAR:2019erd,ALICE:2019onw} have motivated a number of theoretical studies to understand the origin of spin polarization and spin transport of partons in quark gluons plasmas (QGP). It was originally proposed that the large angular momentum created in peripheral collisions could lead to the spin polarization of partons through the spin-orbit interaction and later inherited by the spin polarization of hadrons \cite{Liang:2004ph,Liang:2004xn}. In global thermal equilibrium, the spin polarization of hadrons is dictated by thermal vorticity \cite{Becattini2013a,Fang:2016vpj}. This theoretical description remarkably matches the measurements of global spin polarization of $\Lambda$ hyperons \cite{Karpenko:2016jyx,Li:2017slc,Xie:2017upb,Wei:2018zfb,Ryu:2021lnx}. Nevertheless, the contribution from solely thermal vorticity \cite{Becattini:2017gcx,Xia:2018tes} disagrees with the later measured local spin polarization \cite{STAR:2019erd}, which implies further corrections beyond the global equilibrium condition need to be considered. Several corrections in \cite{Hidaka:2017auj,Liu:2020dxg,Liu:2021uhn,Fu:2021pok,Becattini:2021suc,Becattini:2021iol,Yi:2021ryh,Florkowski:2021xvy} and out of local equilibrium \cite{Hidaka:2017auj,Hidaka:2018ekt,Yang:2018lew,Shi:2020htn,Fang:2022ttm,Wang:2022yli,Weickgenannt:2022zxs,Weickgenannt:2022qvh,Bhadury:2020puc,Bhadury:2020cop,Buzzegoli:2022kqx,Bhadury:2022ulr} have been recently studied, while these contributions mostly come from gradients of hydrodynamic variables such as the thermal-shear correction \cite{Hidaka:2017auj,Liu:2021uhn,Becattini:2021suc} or the chemical-potential gradient \cite{Hidaka:2017auj,Liu:2020dxg}. See Ref.~\cite{Becattini:2022zvf} for a recent review and more references for the spin polarization in heavy ion collisions. 

In addition to the observations of spin polarization of hadrons, there have been further measurements for spin alignment of vector mesons characterized by the deviation of the longitudinal ($00$) component of the spin density matrix $\rho_{00}$ from $1/3$ \cite{ALICE:2019aid,STAR:2022fan,ALICE:2022sli}. As inferred by the spin coalescence model \cite{Liang:2004xn,Yang:2017sdk}, the unexpectedly large deviation observed in experiments implies strong spin polarization of the composite quark and antiquark or more precisely their spin correlation compared with those extracted from the spin polarization of $\Lambda$ hyperons. Moreover, there exist both quantitative and qualitative differences for the spin alignment between distinct flavors and different collision energies. For example, a value $\rho_{00}<1/3$ is observed for both $\phi$ and $K^{*0}$ mesons with small transverse momenta at LHC energies \cite{ALICE:2019aid}, whereas $\rho_{00}>1/3$ for $\phi$ and $\rho_{00}\approx 1/3$ for $K^{*0}$ were found for global spin alignment at RHIC \cite{STAR:2022fan}. 
Various theoretical mechanisms \cite{Sheng:2019kmk,Sheng:2020ghv,Xia:2020tyd,Muller:2021hpe,Yang:2021fea,Goncalves:2021ziy,Sheng:2022wsy,Li:2022neh},beyond the contributions from hydrodynamic gradients, have been proposed as one may not expect that the collision energy and flavor dependence could be simultaneously explained by a single effect. At this time the puzzle remains unsolved. 

In Refs.~\cite{Muller:2021hpe,Yang:2021fea}, using quantum kinetic theory (QKT) for the coupled vector and axial vecotor current evolution of spin-$1/2$ massless \cite{Son:2012wh,Stephanov:2012ki,Chen:2012ca,Hidaka:2016yjf} and massive fermions \cite{Gao:2019znl,Weickgenannt:2019dks,Hattori:2019ahi,Wang:2019moi,Yang:2020hri,Wang:2020pej,Weickgenannt:2020aaf} in phase space (see also \cite{Hidaka:2022dmn} for a recent review and references), two of us proposed that the turbulent color fields from weakly coupled anisotropic QGP could potentially result in $\rho_{00}<1/3$ for spin alignment. As opposed to most studies focusing on late-time effects upon spin polarization, we also pointed out that the dynamical source term in this framework could capture the early-time effects that results in the spin polarization at freeze-out. In general, such early-time effects will be further affected by collisions in late times, from which the spin polarization could be suppressed by relaxation or enhanced by quantum corrections from gradient terms such as vorticity \cite{Kapusta:2019sad,Li:2019qkf,Yang:2020hri,Weickgenannt:2020aaf,Wang:2020pej,Wang:2021qnt,Fang:2022ttm,Hongo:2022izs,Wang:2022yli}. However, the effect from strong background fields $\sim \mathcal{O}(g^2)$ with $g$ the QCD coupling overwhelms the collisional effect from scattering with hard partons $\sim\mathcal{O}(g^4)$ or $\sim\mathcal{O}(g^4\ln g)$ at weak coupling \cite{Li:2019qkf,Yang:2020hri,Hongo:2022izs} as a systematic analysis of the kinetic equations shows. Nevertheless, the magnitudes of such turbulent color fields, originating from Weibel-type instabilities in expanding QGP \cite{Mrowczynski:1988dz,Mrowczynski:1993qm,Romatschke:2003ms}, remains unknown. 

On the other hand, in the early stage of high-energy nuclear collisions before the formation of QGP, the soft gluons with large densities sourced by the hard partons could be described by dynamical color fields encoded in classical Yang-Mills equations and form the highly-dense matter known as glasma \cite{Lappi:2006fp,Lappi:2006hq} in the color glass condensate (CGC) effective theory \cite{McLerran:1993ni,McLerran:1993ka,McLerran:1994vd} (see also \cite{Gelis:2010nm,Albacete:2014fwa} for reviews). It is hence of interest to explore how the strong color fields in the glasma phase could influence the spin polarization and correlation of quarks and antiquarks via the dynamical source term obtained from the QKT \cite{Muller:2021hpe,Yang:2021fea}. Some recent studies have shown a substantial effect from these color fields in glasma on jet quenching of hard probes due to accumulating momentum transfer from the soft gluons even though the glasma phase lasts for only a relatively short period compared with QGP \cite{Carrington:2021dvw,Carrington:2022bnv}. There also exists the study of the glasma influence on angular momentum fluctuations of heavy quarks~\cite{Pooja:2022ojj}. 
One may analogously anticipate the accumulated angular momentum transfer from these soft gluons yields the spin polarization and correlation of quarks and antiquarks traveling through the glasma. In this paper, we investigate 
 such effects by studying the dynamical source term from QKT augmented by the color fields analytically solved from linearized (Abelianized) Yang-Mills equations in Ref.~\cite{Guerrero-Rodriguez:2021ask}, from which the spin 
 polarization and correlation of quarks and antiquarks at small momentum and central rapidity are derived in the integral form of the gluon distribution in the glasma. By  adopting the Golec-Biernat--Wusthoff (GBW) dipole distribution \cite{Golec-Biernat:1998zce}, we numerically estimate the non-vanishing spin correlation out of plane and manifest its enhancement at weak coupling and large collision energy.    

The paper is organized as follows : In Sec.~\ref{sec:general}, we briefly review how dynamical spin polarization of quarks and the related spin correlation associated with spin alignment of vector mesons can be induced by color fields in the framework of QKT. The simplified case at small momentum is 
further examined. In Sec.~\ref{sec:glasma}, we then review the chromo-electric and -magnetic fields derived from linearized Yang-Mills equations in the glasma. In Sec.~\ref{sec:spin_pol_corr}, the dynamical spin polarization 
and spin correlation induced by such color fields in the glasma are accordingly investigated. In Sec.~\ref{sec:GBW}, we further analyze these results with the GBW distribution and make a numerical estimation of the spin 
correlation. Finally, in Sec.~\ref{sec:conclusions}, we present our conclusions and outlook.  Some details of calculations and derivations are presented in appendices.  

Throughout this paper, we use 
the signature $\eta^{\mu\nu} = {\rm diag} (1, -1,-1,-1)  $ of the Minkowski metric and the completely antisymmetric tensor $\epsilon^{\mu\nu\rho\lambda}$ with $\epsilon^{0123}= \epsilon^{txyz}= 1$, where $0,1,2,3$ and $t,x,y,z$ will be used as the space-time indices interchangeably unless specified. We introduce the notations $A^{(\mu}B^{\nu)}\equiv A^{\mu}B^{\nu}+A^{\nu}B^{\mu}$ 
and $A^{[\mu}B^{\nu]}\equiv A^{\mu}B^{\nu}-A^{\nu}B^{\mu}$. We also define the dual field strength of color fields via $\tilde{F}^{a\mu\nu}\equiv\epsilon^{\mu\nu\alpha\beta}F^a_{\alpha\beta}/2$ with $a$ being color indices.

\section{Dynamical spin polarization from quantum kinetic theory}\label{sec:general}

In this section, we briefly summarize the spin polarization and correlation obtained from the framework of QKT and Wigner functions of quarks under background color fields previously derived in Refs.~\cite{Muller:2021hpe,Yang:2021fea}. We also review the connection between spin correlation and spin alignment of vector mesons with an update on the coalescence model. The dynamical spin polarization led by color fields at the small-momentum limit with a more generic form than those in Refs.~\cite{Muller:2021hpe,Yang:2021fea} are presented, which will be later utilized to study the effects from glasma in subsequent sections.
 
\subsection{From spin correlation to spin alignment by color fields}
As shown in Refs.~\cite{Muller:2021hpe,Yang:2021fea}, the spin-polarization spectrum of a quark is described by the spin Cooper-Frye formula \cite{Becattini2013a,Fang:2016vpj}, 
\begin{equation}\label{Spin_CooperFrye}
	\mathcal{P}^{\mu}({\bm p})=\frac{\int d\Sigma\cdot p\mathcal{J}_{5}^{s\mu}(\bm p,X)}{2m\int d\Sigma_{\mu}\mathcal{N}^{s\mu}(\bm p,X)},
\end{equation}
where $\mathcal{J}_{5}^{s\mu}(\bm p,X)$ and $\mathcal{N}^{s\mu}(\bm p,X)$ with $\bm p$ being the quark momentum denote the onshell color-singlet axial-charge-current density and number-current density in phase space, respectively. Here $d\Sigma_{\mu}$ is the normal vector of a freeze-out hyper surface and $m$ is the mass of quarks. Also, $\mathcal{J}_{5}^{s\mu}(\bm p,X)$ and $\mathcal{N}^{s\mu}(\bm p,X)$ are associated with the axial-vector and vector components of the color-singlet Wigner functions of quarks, which can be expressed in terms of the effective spin four-vector $\tilde{a}^{s\mu}(p,X)$ and the vector-charge distribution function $f^{s}_{V}(p,X)$, where the former delineates the dynamical spin evolution and the later describes the energy and charge transport of quarks. Their dynamics are governed by an axial kinetic equation and a scalar kinetic equation as the traditional Boltzmann equation, while the collisional effects are neglected due to the suppression compared to relatively strong background fields at weak coupling as our assumption. In this setup, the leading-order $\mathcal{J}_{5}^{s\mu}(\bm p,X)$ and $\mathcal{N}^{s\mu}(\bm p,X)$ in the $\hbar$ expansion are given by \cite{Muller:2021hpe,Yang:2021fea} 
\begin{eqnarray}
	\mathcal{N}^{s\mu}(\bm p,X)&=& \big(p^{\mu}f^s_{V}\big)_{p_0=\epsilon_{\bm p}},
	\\\label{eq:J5smu_density}
	\mathcal{J}_5^{s\mu}(\bm p,X)&=&\big(\tilde{a}^{ s\mu}+\hbar\tilde{C}_2\mathcal{A}^{\mu}_{Q}\big)_{p_0=\epsilon_{\bm p}},
\end{eqnarray}
where $\epsilon_{\bm p}=\sqrt{|\bm p|^2+m^2}$ is the onshell energy and $\tilde{C}_2=g^2/(2N_c)$ with $N_c$ the number of colors. The $\hbar$ is an expansion parameter characterizing the order of gradient expansion for Wigner functions in phase space. Here the non-dynamical source term, $\mathcal{A}^{\mu}_{Q}$, can be decomposed into
\footnote{The integration of $\epsilon_{\bm p}^{-1}\mathcal{A}^{\mu}_{Q}|_{p_0=\epsilon_{\bm p}}$ over $\bm p$ contributes to an axial-charge current. Here $\epsilon_{\bm p}^{-1}\mathcal{A}^{\mu}_{Q2}$ is a total-derivative term in momentum, which accordingly leads to a vanishing contribution to the axial-charge current. However, it still gives rise to a non-vanishing contribution to the spin-polarization spectrum.}
\begin{eqnarray}\label{eq:AQ}
\mathcal{A}^{\mu}_{Q}|_{p_0=\epsilon_{\bm p}}=\mathcal{A}^{\mu}_{Q1}+\mathcal{A}^{\mu}_{Q2}
\end{eqnarray}
with the explicit form of $\mathcal{A}^{\mu}_{Q1}$ and $\mathcal{A}^{\mu}_{Q2}$ as
\begin{equation}\label{AQmu_origin}
	\mathcal{A}^{\mu}_{Q1}=\bigg[\frac{\partial_{p\kappa}}{2}\int^{p,X}_{k,X'}p^{\beta}\big( \tilde{F}^{a\mu\kappa}(X)F^a_{\alpha\beta}(X')\big)\partial_{p}^{\alpha}f^{\rm s}_V(p,X')\bigg]_{p_0=\epsilon_{\bm p}},
\end{equation}
and
\begin{eqnarray}\nonumber
		\mathcal{A}^{\mu}_{Q2}&=&-\frac{\epsilon_{\bm p}}{2}\partial_{p_\perp\kappa}\bigg[\int^{p,X}_{k,X'}\hat{p}^{\beta}\big( \tilde{F}^{a\mu\kappa}(X)F^a_{\alpha\beta}(X')\big)\partial_{p}^{\alpha}f^{\rm s}_V(p,X')\bigg]_{p_0=\epsilon_{\bm p}}
		\\
		&=&\frac{1}{2\epsilon_{\bm p}^2}(p_{\perp\kappa}-\epsilon_{\bm p}^2\partial_{p_{\perp}\kappa})\bigg[\int^{p,X}_{k,X'}p^{\beta}\big( \tilde{F}^{a\mu\kappa}(X)F^a_{\alpha\beta}(X')\big) \partial_{p}^{\alpha}f^{\rm s}_V(p,X')\bigg]_{p_0=\epsilon_{\bm p}},
\end{eqnarray}
where $F^a_{\alpha\beta}$ denotes the color gauge field in the glasma and $\hat{p}_{\mu}\equiv p_{\mu}/p_0$. Furthermore, we introduced the abbreviation 
\begin{eqnarray}
	\int^{p,X}_{k,X'}\equiv \int d^4k\int\frac{d^4X'}{(2\pi)^4}e^{ik\cdot(X'-X)}\big(\pi\delta(p\cdot k)+iPV(1/p\cdot k)\big).
\end{eqnarray}
Throughout this section, $V^{\mu}_{\perp}$ represents the spatial component of any four-vector $V^{\mu}$. The notation $PV(x)$ denotes the principal value of $x$. Note that the integral $\int^{p,X}_{k,X'}$ integrates over $k$ and $X'$ and leads to a function depending on $p$ and $X$. The notation $\big( F^a_{\kappa\nu}(X)F^a_{\alpha\lambda}(X')\big)$ is understood to implicitly incorporate a gauge link between $X'$ and $X$ for gauge invariance. We will neglect the gauge link in perturbative calculations because it contributes at higher order in the gauge coupling. $F^a_{\kappa\nu}(X)$ and $F^a_{\alpha\lambda}(X')$ here should be still regarded as field operators and thus $\mathcal{P}^{\mu}({\bm p})$ in Eq.~(\ref{Spin_CooperFrye}) is also a Fock space operator. This means that one has to further take the ensemble average of the correlation function for color fields, $\langle F^a_{\kappa\nu}(X)F^a_{\alpha\lambda}(X')\rangle$, to evaluate the spin polarization spectrum $\langle\mathcal{P}^{\mu}({\bm p})\rangle$ when comparing to the experimental observable.

On the other hand, we have to also include the contribution from $\tilde{a}^{s\mu}$ for dynamical spin polarization. The dynamics of $\tilde{a}^{s\mu}$ is governed by the color-singlet axial kinetic equation,
\begin{eqnarray}\label{AKE_signlet_simplify}
	0=\delta(p^2-m^2)\Big(p\cdot\partial\tilde{a}^{\rm s\mu}(p,X)-\partial_{p}^{\kappa}\mathscr{D}_{\kappa}[\tilde{a}^{s\mu}]
	+\hbar\partial_{p}^{\kappa}\big(\mathscr{A}^{\mu}_{\kappa}[f^{\rm s}_{V}]\big)\Big),
\end{eqnarray}
where 
\begin{eqnarray}
	\mathscr{D}_{\kappa}[O]=\tilde{C}_2\int^{p,X}_{k,X'}p^{\lambda}p^{\rho}\big(F^a_{\kappa\lambda}(X)
	F^a_{\alpha\rho}(X')\big) \partial_{p}^{\alpha}O(p,X')
\end{eqnarray}
and
\begin{eqnarray}
	\mathscr{A}^{\mu}_{\kappa}[O]=\frac{\tilde{C}_2}{2}
	\epsilon^{\mu\nu\rho\sigma}\int^{p,X}_{k,X'}p^{\lambda}p_{\rho}\Big(\partial_{X'\sigma}\big( F^a_{\kappa\lambda}(X)F^a_{\alpha\nu}(X')\big)
	+\partial_{X\sigma}\big( F^a_{\kappa\nu}(X)F^a_{\alpha\lambda}(X')\big)\Big)\partial^{\alpha}_{p}O(p,X').
\end{eqnarray}
At weak coupling, given no initial polarization, $\tilde{a}^{s\mu}$ has to be induced by the dynamical source term $\hbar\partial_{p}^{\kappa}\big(\mathscr{A}^{\mu}_{\kappa}[f^{\rm s}_{V}]\big)$ in Eq.~(\ref{AKE_signlet_simplify}), which yields $\tilde{a}^{s\mu}\sim \mathcal{O}(g^2)$ and the diffusion term $\partial_{p}^{\kappa}\mathscr{D}_{\kappa}[\tilde{a}^{s\mu}]$ is accordingly of $\mathcal{O}(g^4)$. Unless we consider the evolution for sufficiently long time, one may assume $p\cdot\partial\tilde{a}^{\rm s\mu}(p,X)\gg \partial_{p}^{\kappa}\mathscr{D}_{\kappa}[\tilde{a}^{s\mu}]$.
To obtain the solution of $\tilde{a}^{s\mu}$ from the kinetic equation, we may utilize the solution,
\begin{eqnarray}\label{eq:useful_int}\nonumber
	\tilde{a}^{\mu}(p,X)&=&\int^{p,X}_{k,X'}G^{\mu}(p,X')
	\\
	&=&\int\frac{d \delta X_0}{2p_0}\Big(1+{\rm sgn}(\delta X_0)\Big)G^{\mu}(p,X')|_{\delta X_{x,y}=0,\delta X_z=p_z\delta X_0/p_0},
\end{eqnarray}
for the differential equation,
\begin{eqnarray}\label{eq:ex_diffeq}
p\cdot\partial \tilde{a}^{\mu}(p,X)=G^{\mu}(p,X),
\end{eqnarray}
with an arbitrary function $G^{\mu}(p,X)$ independent of $\tilde{a}^{\mu}(p,X)$, where $\delta X=X-X'$.
By neglecting the diffusion term, one finds 
\begin{eqnarray}
\tilde{a}^{s\mu}(p,X)=-\frac{\hbar}{2p_0}\int d\delta X_0\Big(1+{\rm sgn}(\delta X_0)\Big)\partial_{p}^{\kappa}\big(\mathscr{A}^{\mu}_{\kappa}[f^{\rm s}_{V}](X')\big)|_{\delta X_{x,y}=0,\delta X_z=p_z\delta X_0/p_0}.
\end{eqnarray}
For notational convenience, we may write 
\begin{eqnarray}\nonumber \label{eq:a_sol}
	\tilde{a}^{s\mu}(p,X)&=&-\hbar\int^{p,X}_{k,X'}\partial_{p}^{\kappa}\big(\mathscr{A}^{\mu}_{\kappa}[f^{\rm s}_{V}](X')\big)
	\\\nonumber
	&=&-\frac{\hbar\tilde{C}_2}{2}
	\epsilon^{\mu\nu\rho\sigma}\int^{p,X}_{k,X'}\partial_{p}^{\kappa}\int^{p,X'}_{k',X''}p^{\lambda}p_{\rho}\Big(\partial_{X''\sigma}\big( F^a_{\kappa\lambda}(X')F^a_{\alpha\nu}(X'')\big)
	+\partial_{X'\sigma}\big( F^a_{\kappa\nu}(X')F^a_{\alpha\lambda}(X'')\big)\Big)
	\\
	&&\times\partial^{\alpha}_{p}f^{\rm s}_{V}(p,X'').
\end{eqnarray}
Similar to the non-dynamical source term, the field strengths above are also operators and the ensemble average needs to be taken in the end.

Now, given an explicit expression of spin polarization for a quark $\mathcal{P}_{q}^{\mu}({\bm p})$ and for an antiquark $\mathcal{P}_{\bar{q}}^{\mu}({\bm p})$, one could further compute the ensemble average of spin correlation,
\begin{equation}
	\langle\mathcal{P}_q^{\mu}({\bm p})\mathcal{P}_{\bar{q}}^{\mu}({\bm p})\rangle =\frac{\int d\Sigma_X\cdot p\int d\Sigma_Y\cdot p\langle\mathcal{J}_{5}^{s\mu}(\bm p,X)\mathcal{J}_{5}^{s\mu}(\bm p,Y)\rangle}{4m^2\Big(\int d\Sigma_X\cdot\mathcal{N}^{s}(\bm p,X)\Big)^2}.
\end{equation}
Note that $\langle\mathcal{P}_{q}^{\mu}({\bm p})\mathcal{P}_{\bar{q}}^{\mu}({\bm p})\rangle$ needs not be equal to $\langle\mathcal{P}_{q}^{\mu}({\bm p})\rangle\langle\mathcal{P}_{\bar{q}}^{\mu}({\bm p})\rangle$. From the spin-dependent coalescence model, one may evaluate the $00$ component of the spin density matrix for spin-$1$ vector mesons through \cite{Liang:2004xn,Yang:2017sdk}
\begin{eqnarray}\label{eq:rho00_origin}
	\rho_{00}=\frac{1-\langle\mathcal{P}^{i}_{q}\mathcal{P}^{i}_{\bar{q}}\rangle}{3+\langle\mathcal{P}^{i}_{q}\mathcal{P}^{i}_{\bar{q}}\rangle},
\end{eqnarray}
where the superscript $i$ is the assigned spin quantization axis determined by experimental setup, which may be chosen as the $y$ axis along the direction perpendicular to the reaction plane in heavy ion collisions. The deviation of $\rho_{00}$ from $1/3$ thus implies non-vanishing spin correlation in QGP even when the spin polarization of a quark or an antiquark vanishes. However, there exists a caveat for Eq.~(\ref{eq:rho00_origin}), which is in fact derived based on the assumption that the spin of a quark and of an antiquark are fully polarized along the quantization axis. In a more generic case, as shown in Ref.~\cite{Sheng:2022ffb} (see also appendix \ref{app:density_matrix} for a consistent derivation), Eq.~(\ref{eq:rho00_origin}) should be modified as  
\begin{eqnarray}\label{eq:rho00_modified}
	\rho_{00}\approx \frac{1+\sum_{j=x,y,z}\langle\mathcal{P}^{j}_{q}\mathcal{P}^{j}_{\bar{q}}\rangle-2\langle\mathcal{P}^{i}_{q}\mathcal{P}^{i}_{\bar{q}}\rangle}{3+\sum_{j=x,y,z}\langle\mathcal{P}^{j}_{q}\mathcal{P}^{j}_{\bar{q}}\rangle},
\end{eqnarray}
where $i$ here again denotes the spin quantization axis yet the spin correlations for all other directions need to be incorporated. The crucial difference is that $\rho_{00}=1/3$ even when $\langle\mathcal{P}^{j}_{q}\mathcal{P}^{j}_{\bar{q}}\rangle\neq 0$ is isotropic (i.e. $\langle\mathcal{P}^{x}_{q}\mathcal{P}^{x}_{\bar{q}}\rangle=\langle\mathcal{P}^{y}_{q}\mathcal{P}^{y}_{\bar{q}}\rangle=\langle\mathcal{P}^{z}_{q}\mathcal{P}^{z}_{\bar{q}}\rangle$).
When evaluating $\rho_{00}$ or $\langle\mathcal{P}^{i}_{q}\mathcal{P}^{i}_{\bar{q}}\rangle$ in the glasma phase, instead of simply taking the ensemble average of two chromo-electromagnetic fields from dynamical source terms, we will consider the ensemble average of four field-strength operators therein.

Notably, when comparing the spin polarization led by the non-dynamical source term in Eq.~(\ref{eq:AQ}) and by the dynamical one in Eq.~(\ref{eq:a_sol}), it is found that the former depends on only the color-field correlator at the late time when spin freezes out. On the contrary, the latter is contributed by integrating the color-field correlator over a whole period before the spin freeze-out. Considering the spin polarization from the strong color fields in the glasma phase in early times, which decay shortly after collisions, the dynamical spin polarization should dominate over the non-dynamical one. 
Accordingly, we shall evaluate the spin correlation via
\begin{eqnarray}\nonumber
\langle\mathcal{P}^{i}_{q}\mathcal{P}^{i}_{\bar{q}}\rangle &\sim& \langle \tilde{a}^{si}(p,X)\tilde{a}^{si}(p,Y)\rangle 
\\\nonumber
&=&\Big\langle\int^{p,X}_{k,X'}\partial_{p}^{\kappa}\big(\mathscr{A}^{i}_{\kappa}[f^{\rm s}_{V}](X')\big)\int^{p,Y}_{k,Y'}\partial_{p}^{\kappa}\big(\mathscr{A}^{i}_{\kappa}[f^{\rm s}_{V}](Y')\big)\Big\rangle
	\\\nonumber
&=&\frac{\tilde{C}_2^2}{4}\epsilon^{i\nu\rho\sigma}\epsilon^{i\nu'\rho'\sigma'}\Big\langle\int^{p,X}_{k,X'}\partial_{p}^{\kappa}\int^{p,X'}_{k',X''}p^{\lambda}p_{\rho}\Big(\partial_{X''\sigma}\big( F^a_{\kappa\lambda}(X')F^a_{\alpha\nu}(X'')\big)
+\partial_{X'\sigma}\big( F^a_{\kappa\nu}(X')F^a_{\alpha\lambda}(X'')\big)\Big)
\\\nonumber
&&\times\partial^{\alpha}_{p}f^{\rm s}_{V}(p,X'')
\int^{p,Y}_{\bar{k},Y'}\partial_{p}^{\kappa'}\int^{p,Y'}_{\bar{k}',Y''}p^{\lambda'}p_{\rho'}\Big(\partial_{Y''\sigma'}\big( F^b_{\kappa'\lambda'}(Y')F^b_{\alpha'\nu'}(Y'')\big)
+\partial_{Y'\sigma'}\big( F^b_{\kappa'\nu'}(Y')F^b_{\alpha'\lambda'}(Y'')\big)\Big)
\\
&&\times\partial^{\alpha'}_{p}f^{\rm s}_{V}(p,Y'')\Big\rangle
\end{eqnarray}
from Eq.~(\ref{eq:a_sol}).
Despite the generic $\rho_{00}$ depends on $\langle\mathcal{P}^{j}_{q}\mathcal{P}^{j}_{\bar{q}}\rangle$ for $j=x,y,z$, we will only focus on $\langle\mathcal{P}^{y}_{q}\mathcal{P}^{y}_{\bar{q}}\rangle$ in this paper for simplicity.

\subsection{Simplification in the small-momentum limit}
For practical applications, one has to convert the field strengths into chromo-electric and -magnetic fields, which are explicitly given by
\begin{eqnarray}
	F^{a}_{\mu\nu}=-\epsilon_{\mu\nu\alpha\beta}B^{a\alpha}\bar{n}^{\beta}+E^a_{[\mu}\bar{n}_{\nu]},\quad
	\tilde{F}^{a\mu\nu}=B^{a[\mu}\bar{n}^{\nu]}+\epsilon^{\mu\nu\alpha\beta}E^a_{\alpha}\bar{n}_{\beta},
\end{eqnarray} 
where $\bar{n}^{\mu}=(1,\bm 0)$ denotes the temporal direction.
We will make further approximations to simplify the dynamical source term. Nevertheless, unlike the derivation in Refs.~\cite{Muller:2021hpe,Yang:2021fea}, we will not impose the hierarchy between chromo-electric and -magnetic fields and keep the operator form of color-field correlations for generality. We first assume $f^{s}_V=f_V(p_0)$ as a function only depending on $p_0$ for simplification.
Considering 
\begin{align}
	\mathscr{A}^{\mu}_{\kappa}[f_V(p_0)]=-\frac{\tilde{C}_2}{2}
	\epsilon^{\mu\nu\rho\sigma}\int^{p,X}_{k,X'}p^{\lambda}p_{\rho}\Big(\partial_{X'\sigma}\left( F^a_{\kappa\lambda}(X)E^a_{\nu}(X')\right)
	+\partial_{X\sigma}\left( F^a_{\kappa\nu}(X)E^a_{\lambda}(X')\right)\Big)\partial_{p0}f_V(p_0), \label{AKE_source_term_cont_}
\end{align}
where we omit the $X$ dependence of $\mathscr{A}^{\mu}_{\kappa}[f_V(p_0)]$ for notational convenience,
and using the relation
\begin{eqnarray}\nonumber\label{eq:DkintG}
	\partial_{p\kappa}\int^{p,X}_{k,X'}G(X,\delta X)&=&-\frac{\delta^{0}_{\kappa}}{p_0}\int^{p,X}_{k,X'}G(X,\delta X)
	+\int^{p,X}_{k,X'}\partial_{p\kappa}G(X,\delta X)
	\\
	&&+\int^{p,X}_{k,X'}\delta X_0\frac{\partial G(X,\delta X)}{\partial \delta X^{\mu}_{\perp}}\partial_{p\kappa}\left(\frac{p_{\perp}^{\mu}}{p_0}\right),
\end{eqnarray}
we derive
\begin{eqnarray}\nonumber
	\partial_{p}^{\kappa}\mathscr{A}^{\mu}_{\kappa}[f_V(p_0)]&=&-\frac{1}{p_0}\int^{p,X}_{k,X'}\hat{\mathscr{A}}^{\mu}_{0}[f_V(p_0)]+\int^{p,X}_{k,X'}\partial_{p\kappa}\hat{\mathscr{A}}^{\mu}_{\kappa}[f_V(p_0)]
	+\int^{p,X}_{k,X'}(X_0-X'_0)\frac{\partial \hat{\mathscr{A}}^{\mu}_{\kappa}[f_V(p_0)]}{\partial  X'^{\nu}_{\perp}}
	\\
	&&\times\frac{1}{p_0}(\eta^{\nu\kappa}-\hat{p}^{\nu}\bar{n}^{\kappa}).
\end{eqnarray}
Here we introduce
\begin{eqnarray}
	\hat{\mathscr{A}}^{\mu}_{\kappa}[f_V(p_0)]=-\frac{\tilde{C}_2}{2}
	\epsilon^{\mu\nu\rho\sigma}p^{\lambda}p_{\rho}\Big(\partial_{X'\sigma}\left( F^a_{\kappa\lambda}(X)E^a_{\nu}(X')\right)
	+\partial_{X\sigma}\left( F^a_{\kappa\nu}(X)E^a_{\lambda}(X')\right)\Big)\partial_{p0}f_V(p_0)
\end{eqnarray}
and hence 
\begin{eqnarray}\nonumber
	\partial^{\kappa}_{p}\hat{\mathscr{A}}^{\mu}_{\kappa}[f_V(p_0)]&=&-\frac{\tilde{C}_2}{2}
	\epsilon^{\mu\nu\rho\sigma}\big(\eta^{\kappa\lambda} p_{\rho}+p^{\lambda}\delta^{\kappa}_{\rho}+p^{\lambda}p_{\rho}\delta^{\kappa}_0\partial_{p0}\big)\Big(\partial_{X'\sigma}\left( F^a_{\kappa\lambda}(X)E^a_{\nu}(X')\right)
	\\
	&&+\partial_{X\sigma}\left( F^a_{\kappa\nu}(X)E^a_{\lambda}(X')\right)\Big)\partial_{p0}f_V(p_0).
\end{eqnarray}

To proceed, we postulate a semi-thermal distribution $f_{V}(p_0)=1/(e^{p_0/\Lambda}+1)$ as the distribution function for quarks or antiquarks created in early times of the glasma phase with $\Lambda$ an effective energy scale that should eventually evolve to temperature of the QGP after thermalization. For simplicity, we will also ignore the time dependence of $\Lambda$. By considering low-energy particles with $p_0\ll \Lambda$ such that $f_{V}(p_0)$ is less suppressed, one finds $|\partial_{p_0}f_V(p_0)|\gg |p^{\mu}||\partial^2_{p_0}f_V(p_0)|$. Considering the most relevant regime that substantial spin alignment is observed in experiments, we will further focus on the small-momentum limit such that $\hat{p}^{\mu}_{\perp}\equiv p^{\mu}_{\perp}/p_0\ll 1$ for $p_0=\epsilon_{\bm p}\equiv\sqrt{\bm p^2+m^2}$ being onshell. Consequently, we can approximate
\begin{eqnarray}\label{eq:hatA_small_p}
	\hat{\mathscr{A}}^{\mu}_{\kappa}[f_V(p_0)]\approx \frac{\tilde{C}_2}{2}
	\epsilon^{\mu\nu\rho\sigma}p_0^2\bar{n}_{\rho}\partial_{X'\sigma}\left( E^a_{\kappa}(X)E^a_{\nu}(X')\right)
	\partial_{p0}f_V(p_0),
\end{eqnarray}
and
\begin{eqnarray}
	\partial^{\kappa}_{p}\hat{\mathscr{A}}^{\mu}_{\kappa}[f_V(p_0)]\approx-\frac{\tilde{C}_2}{2}
	p_0\Big(\epsilon^{\mu\nu\rho\sigma}\partial_{X'\sigma}\left( E^a_{\rho}(X)E^a_{\nu}(X')\right)
	+\partial_{X\sigma}\left( B^{a[\mu}(X)E^{a\sigma]}(X')\right)\Big)\partial_{p0}f_V(p_0),
\end{eqnarray}
by dropping the higher-order terms of $O(|\hat{p}^{\mu}_{\perp}|)$. From Eq.~(\ref{eq:hatA_small_p}), one also obtains $\hat{\mathscr{A}}^{\mu}_{0}[f_V(p_0)]/p_0\approx 0$ and
\begin{eqnarray}
	\frac{\partial \hat{\mathscr{A}}^{\mu}_{\kappa}[f_V(p_0)]}{p_0\partial  X'^{\kappa}_{\perp}}
	\approx \frac{\tilde{C}_2}{2}
	\epsilon^{\mu\nu\rho\sigma}p_0\bar{n}_{\rho}\partial^{\kappa}_{X'}\partial_{X'\sigma}\left( E^a_{\kappa}(X)E^a_{\nu}(X')\right)
	\partial_{p0}f_V(p_0).
\end{eqnarray}
It turns out that 
\begin{eqnarray}\nonumber
	\partial_{p}^{\kappa}\mathscr{A}^{\mu}_{\kappa}[f_V(p_0)]&\approx&-
	\frac{\tilde{C}_2}{2}
	p_0\big(\partial_{p0}f_V(p_0)\big)\int^{p,X}_{k,X'}\Big[\epsilon^{\mu\nu\rho\sigma}\partial_{X'\sigma}\left( E^a_{\rho}(X)E^a_{\nu}(X')\right)
	+\partial_{X\sigma}\left( B^{a[\mu}(X)E^{a\sigma]}(X')\right)
	\\
	&&+(X_0'-X_0)\epsilon^{\mu\nu\rho\sigma}\bar{n}_{\rho}\partial^{\kappa}_{X'}\partial_{X'\sigma}\left( E^a_{\kappa}(X)E^a_{\nu}(X')\right)
	\Big]
\end{eqnarray}
as the leading-order contribution in the small-momentum limit. One thus arrives at
\begin{eqnarray}\nonumber
	\tilde{a}^{s\mu}(p,X)&=&\frac{\tilde{C}_2}{2}
	p_0\big(\partial_{p0}f_V(p_0)\big)\int^{p,X}_{k,X'}\int^{p,X'}_{k',X''}\Big[\epsilon^{\mu\nu\rho\sigma}\partial_{X''\sigma}\left( E^a_{\rho}(X')E^a_{\nu}(X'')\right)
	+\partial_{X'\sigma}\left( B^{a[\mu}(X')E^{a\sigma]}(X'')\right)
	\\
	&&+(X_0''-X_0')\epsilon^{\mu\nu\rho\sigma}\bar{n}_{\rho}\partial^{\kappa}_{X''}\partial_{X''\sigma}\left( E^a_{\kappa}(X')E^a_{\nu}(X'')\right)
	\Big].
\end{eqnarray}
When choosing $\mu=x,y,z$ (or equivalently  $\mu=1,2,3$ ), each component of $\tilde{a}^{s\mu}(p,X)$ explicitly reads
\begin{eqnarray}\nonumber
	\tilde{a}^{sx}(p,X)
	&=&-\Bigg(\frac{\tilde{C}_2}{2}
	p_0\big(\partial_{p0}f_V(p_0)\big)\Bigg)\int^{p,X}_{k,X'}\int^{p,X'}_{k',X''}\Big[\partial_{X''0}\left(E_{[3}^a(X') E_{2]}^a (X'')\right)
	+\partial^{\nu}_{X'}\left( B^{a}_{[1}(X')E^{a}_{\nu]}(X'')\right) \\\label{eq:exp_asx_origin}
	&&
	+(X_0''-X_0')\partial_{X''}^{\nu}\partial_{X''[2}\left(E_{\nu}^a(X') E_{3]}^a (X'')\right)\Big],
\end{eqnarray}
\begin{eqnarray}\nonumber
	\tilde{a}^{sy}(p,X)
	&=&-\Bigg(\frac{\tilde{C}_2}{2}
	p_0\big(\partial_{p0}f_V(p_0)\big)\Bigg)\int^{p,X}_{k,X'}\int^{p,X'}_{k',X''}\Big[\partial_{X''0}\left(E_{[1}^a(X') E_{3]}^a (X'')\right)
	+\partial^{\nu}_{X'}\left( B^{a}_{[2}(X')E^{a}_{\nu]}(X'')\right) \\\label{eq:exp_asy_origin}
	&&
	+(X_0''-X_0')\partial_{X''}^{\nu}\partial_{X''[3}\left( E_{\nu}^a(X') E_{1]}^a (X'')\right)\Big],
\end{eqnarray}
and
\begin{eqnarray}\nonumber
	\tilde{a}^{sz}(p,X)
	&=&-\Bigg(\frac{\tilde{C}_2}{2}
	p_0\big(\partial_{p0}f_V(p_0)\big)\Bigg)\int^{p,X}_{k,X'}\int^{p,X'}_{k',X''}\Big[\partial_{X''0}\left( E_{[2}^a(X') E_{1]}^a (X'')\right)
	+\partial^{\nu}_{X'}\left( B^{a}_{[3}(X')E^{a}_{\nu]}(X'')\right) \\	\label{eq:exp_asz_origin}
	&&
	+(X_0''-X_0')\partial_{X''}^{\nu}\partial_{X''[1}\left( E_{\nu}^a(X') E_{2]}^a (X'')\right)\Big],
\end{eqnarray}
which are the building blocks of the follow-up computations.

\section{Color fields in the glasma}\label{sec:glasma}
In the infinite-momentum frame for the parton content of a nucleus described by the CGC framework \cite{McLerran:1993ni,McLerran:1993ka,McLerran:1994vd}, the dynamics of the soft modes (small-$x$) of gluons with large occupation numbers follows the classical Yang-Mills equation,
\begin{eqnarray}\label{eq:YM_eq}
[D_{\mu},F^{\mu\nu}]=J^{\nu},
\end{eqnarray}  
where $D_{\mu}$ is the covariant derivative augmented by the non-Abelian gauge field and $F^{\mu\nu}$ is the gluonic field strength. Here we omit the color indices for brevity. In addition, $J^{\mu}$ serves as a source term coming from valence quarks as the hard modes, for which the explicit expressions in term of color-source densities for collisions of two nuclei and further analyses can be found in Ref.~\cite{Kovner:1995ts}. Generically, Eq.~(\ref{eq:YM_eq}) may only be solved numerically with prescribed initial conditions \cite{Krasnitz:1998ns,Krasnitz:2001qu,Schenke:2012hg,Schenke:2012wb,Mantysaari:2017cni,Jia:2020sbn}, while analytic approximations can be obtained in either the weak-field limit of the dilute-dense system \cite{Kovner:1995ts,Kovchegov:1997ke,Dumitru:2001ux,McLerran:2016snu,Lappi:2017skr} or the early-time limit via the perturbative expansion in an infinitesimal proper time \cite{Fries:2006pv,Fujii:2008km,Chen:2015wia,Carrington:2020ssh}. However, for our purpose to study the memory effect of dynamical spin polarization, the inclusion of late-time dynamics for color fields is indispensable. As a result, we will instead adopt the linearized (Abelianized) approximation in Ref.~\cite{Guerrero-Rodriguez:2021ask}, from which the late-time dynamics can be captured by sacrificing gauge invariance and presuming the nonlinear corrections upon quantities of our interest are small.          
According to Ref.~\cite{Guerrero-Rodriguez:2021ask}, based on the definitions $E^{\eta}\equiv F_{\tau\eta}/\tau$, $E^i_T\equiv F_{\tau i}$, $B^{\eta}\equiv-\epsilon^{ij}F_{ij}/2$, and $B^i\equiv -\epsilon^{ij}F_{j\tau}/\tau$ in Milne coordinates with $\tau$ the proper time and $\eta$ spacetime rapidity, the chromo-electric and -magnetic fields in the glasma are found to be 
\begin{align}\label{eq:EM_CGC_E3}
	E^{\eta}(\tau,x_{\perp})&=-ig\delta^{ij}\!\!\int \!\frac{d^2q_{\perp}}{(2\pi)^2}\!\!\int \!d^2u_{\perp}[\alpha^{i}_{1}(u_{\perp}),\alpha^{j}_{2}(u_{\perp})]
	\times\!J_0(q\tau)e^{iq_{\perp}(x-u)_{\perp}},
\\
	B^{\eta}(\tau,x_{\perp})&=-ig\epsilon^{ij}\!\!\int \!\frac{d^2q_{\perp}}{(2\pi)^2}\!\!\int \!d^2u_{\perp}[\alpha^{i}_{1}(u_{\perp}),\alpha^{j}_{2}(u_{\perp})]
	\times\!J_0(q\tau)e^{iq_{\perp}(x-u)_{\perp}},
\\
E^{i}_{T}(\tau,x_{\perp})&=-g\epsilon^{ij}\epsilon^{kl}\int \!\frac{d^2q_{\perp}}{(2\pi)^2}\frac{q^j}{q}\!\!\int \!d^2u_{\perp}[\alpha^{k}_{1}(u_{\perp}),\alpha^{l}_{2}(u_{\perp})]
\times\!J_1(q\tau)e^{iq_{\perp}(x-u)_{\perp}},
\\\label{eq:EM_CGC_Bi}
B^{i}_{T}(\tau,x_{\perp})&=-g\epsilon^{ij}\delta^{kl}\int \!\frac{d^2q_{\perp}}{(2\pi)^2}\frac{q^j}{q}\!\!\int \!d^2u_{\perp}[\alpha^{k}_{1}(u_{\perp}),\alpha^{l}_{2}(u_{\perp})]
\times\!J_1(q\tau)e^{iq_{\perp}(x-u)_{\perp}},
\end{align}
which yield
\begin{align}
	E^{c\eta}(\tau,x_{\perp})&=gf^{abc}\delta^{ij}\!\!\int \!\frac{d^2q_{\perp}}{(2\pi)^2}\!\!\int \!d^2u_{\perp}\alpha^{i,a}_{1}(u_{\perp})\alpha^{j,b}_{2}(u_{\perp})
	\times\!J_0(q\tau)e^{iq_{\perp}(x-u)_{\perp}},
	\\
	B^{c\eta}(\tau,x_{\perp})&=gf^{abc}\epsilon^{ij}\!\!\int \!\frac{d^2q_{\perp}}{(2\pi)^2}\!\!\int \!d^2u_{\perp}\alpha^{i,a}_{1}(u_{\perp})\alpha^{j,b}_{2}(u_{\perp})
	\times\!J_0(q\tau)e^{iq_{\perp}(x-u)_{\perp}},
	\\
	E^{ci}_T(\tau,x_{\perp})&=-igf^{abc}\epsilon^{ij}\epsilon^{kl}\int \!\frac{d^2q_{\perp}}{(2\pi)^2}\frac{q^j}{q}\!\!\int \!d^2u_{\perp}\alpha^{k,a}_{1}(u_{\perp})\alpha^{l,b}_{2}(u_{\perp})
	\times\!J_1(q\tau)e^{iq_{\perp}(x-u)_{\perp}},
	\\
	B^{ci}_T(\tau,x_{\perp})&=-igf^{abc}\epsilon^{ij}\delta^{kl}\int \!\frac{d^2q_{\perp}}{(2\pi)^2}\frac{q^j}{q}\!\!\int \!d^2u_{\perp}\alpha^{k,a}_{1}(u_{\perp})\alpha^{l,b}_{2}(u_{\perp})
	\times\!J_1(q\tau)e^{iq_{\perp}(x-u)_{\perp}},
\end{align}
where $f^{abc}$ represents the structure constant, $J_{\alpha}(x)$ denotes the Bessel function of the first kind, and $\alpha^{i,a}_{1,2}$ correspond to the transverse gauge fields sourced by the color-charge densities from nuclei $1,2$. Note that the indices $i,j,k,l$ above are along the transverse direction with respect to the light-cone orientations. Also, the subscript $\perp$ is used to denotes the component perpendicular to the light cones and $q_{\perp}x_{\perp}\equiv \sum_{i}q^i_{\perp}x^i_{\perp}$. That is, $i=x,y$. We will hereafter adopt this convention for an arbitrary vector with the subscript $\perp$ (not to confuse it with the convention used in the previous section). 

From the transformation of coordinates, $\tau=\sqrt{t^2-z^2}$ and $\eta=\frac{1}{2}\ln\left(\frac{t+z}{t-z}\right)$, one finds
\begin{eqnarray}
E^{z}\equiv F^a_{0z}=F_{\tau\eta}/\tau\equiv E^{\eta},\quad B^{z}\equiv-\frac{\epsilon^{0zjk}}{2}F_{jk}= -\epsilon^{ij}F^a_{ij}/2\equiv B^{\eta}
\end{eqnarray}
for the longitudinal fields, where $\epsilon^{xy}=-\epsilon^{yx}=1$. For the transverse fields, it is defined that $E^i_T\equiv F_{\tau i}$ and $B^i_{T}\equiv -\epsilon^{ij}F_{j\eta}/\tau$. One hence derives 
\begin{eqnarray}
E^y\equiv F_{0y}=\frac{t}{\tau}F_{\tau y}-\frac{z}{\tau^2}F_{\eta y}=\cosh\eta E^y_T-\sinh \eta B^x_T,\quad 
E^x=\cosh\eta E^x_T+\sinh \eta B^y_T,
\end{eqnarray}
and
\begin{eqnarray}
B^y=\epsilon^{xy}F_{xz}=\frac{-\epsilon^{yx}}{\tau^2}\big(tF_{x\eta}-zF_{x\tau}\big)=\cosh\eta B_T^y+\sinh\eta E_T^x,
\quad B^x=\cosh\eta B_T^x-\sinh\eta E_T^y.
\end{eqnarray}
Here we have used 
\begin{eqnarray}
\partial_{t}\tau=\frac{t}{\tau},\quad \partial_{z}\tau=-\frac{z}{\tau},\quad \partial_{t}\eta=-\frac{z}{\tau^2},\quad \partial_{z}\eta=\frac{t}{\tau^2},
\end{eqnarray}
and $t=\tau\cosh\eta$ and $z=\tau\sinh\eta$. 

From  Ref.~\cite{Guerrero-Rodriguez:2021ask}, it turns out that the following correlators vanish,
\begin{eqnarray}
	\langle E_T^{ai}(X')B_T^{a'j}(X'')\rangle&=&0,\quad \langle E^{a\eta}(X')B^{a'\eta}(X'')\rangle=0,    \label{eq:pocorre}\\
	\langle E_T^{ai}(X')E^{a'\eta}(X'')\rangle&=&0, \quad  \langle  B_T^{ai}(X')B^{a'\eta}(X'')\rangle=0,\label{eq:pocorre1}
\end{eqnarray}
based on the correlation of gluon fields from the same nucleus,
\begin{eqnarray}
\langle \alpha_{1,2}^{i,a}(u_{\perp})\alpha_{1,2}^{j,b}(v_{\perp})\rangle=\frac{\delta^{ab}}{2}\Big[\delta^{ij}G_{1,2}(u_{\perp},v_{\perp})+\Big(\delta^{ij}-2\frac{(u-v)^i_{\perp}(u-v)^j_{\perp}}{|u_{\perp}-v_{\perp}|^2}\Big)h_{1,2}(u_{\perp},v_{\perp})\Big],
\end{eqnarray}
and $\langle \alpha_{1}^{i,a}(u_{\perp})\alpha_{2}^{j,b}(v_{\perp})\rangle=\langle \alpha_{2}^{i,a}(u_{\perp})\alpha_{1}^{j,b}(v_{\perp})\rangle=0$. Here $G_{1,2}$ and $h_{1,2}$ correspond to the unpolarized and linearly polarized gluon distribution functions of nuclei $1$ and $2$, respectively.
The non-vanishing correlators can be written as
\begin{eqnarray}
	\langle E_T^{ai}(X')E_T^{a'j}(X'')\rangle&=&-\bar{N}_c\delta^{aa'}\epsilon^{in}\epsilon^{jm}\int^{X'}_{\perp;q,u}\int^{X''}_{\perp;l,v}\Omega_{-}(u_{\perp},v_{\perp})
	\frac{q^n l^m}{ql}
	\times\!J_1(qX'_0)J_1(lX''_0),    \label{eq:eiejcorr_sim}\\
	\langle B_T^{ai}(X')B_T^{a'j}(X'')\rangle&=&-\bar{N}_c\delta^{aa'}\epsilon^{in}\epsilon^{jm}\int^{X'}_{\perp;q,u}\int^{X''}_{\perp;l,v}
	\Omega_{+}(u_{\perp},v_{\perp})
	\frac{q^n l^m}{ql}
	\times\!J_1(qX'_0)J_1(lX''_0),    \label{eq:bibjcorr_sim}\\
	\langle E_T^{ai}(X')B^{a'\eta}(X'')\rangle&=&-i\bar{N}_c\delta^{aa'}\epsilon^{in}\int^{X'}_{\perp;q,u}\int^{X''}_{\perp;l,v}
	\Omega_{-}(u_{\perp},v_{\perp})
	\frac{q^n}{q}
	\times\!J_1(qX'_0)J_0(lX''_0),      
	\label{eq:eib3corr_sim}\\
	\langle B_T^{ai}(X')E^{a'\eta}(X'')\rangle&=&-i\bar{N}_c\delta^{aa'}\epsilon^{in}\int^{X'}_{\perp;q,u}\int^{X''}_{\perp;l,v}
	\Omega_{+}(u_{\perp},v_{\perp})
	\frac{q^n}{q}
	\times\!J_1(qX'_0)J_0(lX''_0),
	\label{eq:bie3corr_sim}\\
	\langle E^{a\eta}(X')E^{a'\eta}(X'')\rangle&=&\bar{N}_c\delta^{aa'}\int^{X'}_{\perp;q,u}\int^{X''}_{\perp;l,v}
	\Omega_{+}(u_{\perp},v_{\perp})
	\times\!J_0(qX'_0)J_0(lX''_0),
	\label{eq:e3e3corr_sim}\\
	\langle B^{a\eta}(X')B^{a'\eta}(X'')\rangle&=&\bar{N}_c\delta^{aa'}\int^{X'}_{\perp;q,u}\int^{X''}_{\perp;l,v}
	\Omega_{-}(u_{\perp},v_{\perp})
	\times\!J_0(qX'_0)J_0(lX''_0),   \label{eq:b3b3corr}
\end{eqnarray}
where
\begin{eqnarray}
	\bar{N}_c\equiv \frac{1}{2}g^2N_c,
\end{eqnarray}
\begin{eqnarray}
	\Omega_{\mp}(u_{\perp},v_{\perp})=\left[G_1(u_{\perp},v_{\perp})G_2(u_{\perp},v_{\perp})\mp h_1(u_{\perp},v_{\perp})h_2(u_{\perp},v_{\perp})\right],
\end{eqnarray}
and
\begin{eqnarray}\nonumber
	\int^{X'}_{\perp;q,u}\equiv\int \!\frac{d^2q_{\perp}}{(2\pi)^2}\int \!d^2u_{\perp}e^{iq_{\perp}(X'-u)_{\perp}}.
\end{eqnarray}
Note that we also have the useful relation such as
\begin{eqnarray}\label{eq:rel_DE_DB}
	\partial_{Xi}E_T^{ai}(X)=\partial_{Xi}B_T^{ai}(X)=0
\end{eqnarray}
from Eqs.~(\ref{eq:EM_CGC_E3})-(\ref{eq:EM_CGC_Bi}). The color-field correlators above will then be employed to evaluate the spin polarization and correlation.

\section{Dynamical spin polarization and correlation from glasma}\label{sec:spin_pol_corr}
This section is dedicated to the derivation of spin polarization and correlation coming from effective spin four vectors in the compact integral form of color-field correlators from glasma or more precisely gluon distribution functions of nuclei introduced in the previous section. The derivation is rather technical and hence only critical steps are presented, while more detailed computations can be found in appendices. In summary, $\langle\tilde{a}^{sx}(p,X)\rangle=\langle\tilde{a}^{sy}(p,X)\rangle=0$ and  $\langle\tilde{a}^{sz}(p,X)\rangle$ is shown in Eq.~(\ref{eq:asz}) with non-vanishing components in Eq.~(\ref{eq:Ab_final}) and Eq.~(\ref{eq:Ac_final}) for spin polarization. For the out-of-plane (perpendicular to the reaction plane) spin correlation, the primary result is shown in Eq.~(\ref{eq:ayays}) with the explicit expressions in Eqs.~(\ref{eq:I1_final}), (\ref{eq:I2_final}), and (\ref{eq:I3_final}).  

\subsection{Spin polarization}\label{sec:spin_pol}
For simplicity and practical reasons, we will focus on the mid-rapidity region for $\eta\rightarrow 0$ such that $E^{x,y}\approx E_T^{x,y}$ and $B^{x,y}\approx B_T^{x,y}$, which allows us to conveniently compute the integrals of color-field correlators in Minkowski coordinates.
We may now first evaluate $\langle \mathcal{P}^y_q\rangle \sim \langle \tilde{a}^{sy}(p,X)\rangle$. Given 
Eq.~(\ref{eq:pocorre}), one obtains
\begin{eqnarray}\nonumber\label{eq:exp_asy}
	\langle\tilde{a}^{sy}(p,X)\rangle
	&=&-\Bigg(\frac{\tilde{C}_2}{2}
	p_0\big(\partial_{p0}f_V(p_0)\big)\Bigg)\int^{p,X}_{k,X'}\int^{p,X'}_{k',X''}\Big[\partial_{X''0}\left\langle E_{[1}^a(X') E_{3]}^a (X'')\right\rangle\\
	&& -\partial_{X'3}\left\langle B^{a}_{[2}(X')E^{a}_{3]}(X'')\right\rangle
	+(X_0''-X_0')\partial^k_{X''}\partial_{X''[3}\left\langle E_k^a(X') E_{1]}^a (X'')\right\rangle\Big],
\end{eqnarray}
where we have assigned $(x,y,z)=(x^1,x^2,x^3)$. Since $\partial_{X'3}\approx \tau^{-1}\partial_{\eta'}$ and $\partial_{X'0}\approx \partial_{\tau'}$ when $\eta\rightarrow 0$ and the color fields are rapidity-independent, Eq.~(\ref{eq:exp_asy}) reduces to
\begin{eqnarray}\nonumber\label{eq:exp_asy2}
	\langle\tilde{a}^{sy}(p,X)\rangle
	&\approx&-\Bigg(\frac{\tilde{C}_2}{2}
	p_0\big(\partial_{p0}f_V(p_0)\big)\Bigg)\int^{p,X}_{k,X'}\int^{p,X'}_{k',X''}\Big[\partial_{X''0}\left\langle E_{[1}^a(X') E_{3]}^a (X'')\right\rangle
	\\
	&&+(X_0''-X_0')\Big(\partial^2_{X''1}\left\langle E_1^a(X') E_3^a (X'')\right\rangle
	+\partial_{X''2}\partial_{X''1}\left\langle E_2^a(X') E_3^a (X'')\right\rangle\Big)\Big].
\end{eqnarray}
According to Eq.~(\ref{eq:pocorre1}), one consequently concludes $\langle\tilde{a}^{sy}(p,X)\rangle=0$ from Eq.~(\ref{eq:exp_asy2}). Similarly, it is found $\langle\tilde{a}^{sx}(p,X)\rangle=0$ by symmetry, while this is not the case for $\langle\tilde{a}^{sz}(p,X)\rangle$. 

The dynamical spin polarization now may be contributed by  
\begin{eqnarray}\label{eq:asz}
\langle\tilde{a}^{sz}(p,X)\rangle\approx -\Bigg(\frac{\tilde{C}_2}{2}
p_0\big(\partial_{p0}f_V(p_0)\big)\Bigg)\big(\mathcal{A}_{a}+\mathcal{A}_{b}+\mathcal{A}_{c}\big),
\end{eqnarray}
where
\begin{eqnarray}
\mathcal{A}_{a}=\int^{p,X}_{k,X'}\int^{p,X'}_{k',X''}\partial_{X''0}\langle E_{[2}^a(X') E_{1]}^a (X'')\rangle,
\end{eqnarray}
\begin{eqnarray}
\mathcal{A}_{b}=+\int^{p,X}_{k,X'}\int^{p,X'}_{k',X''}\big(\partial^{1}_{X'}\langle B^{a}_{[3}(X')E^{a}_{1]}(X'')\rangle
+\partial^{2}_{X'}\langle B^{a}_{[3}(X')E^{a}_{2]}(X'')\rangle\big),
\end{eqnarray} 
and
\begin{eqnarray}
\mathcal{A}_{c}=\int^{p,X}_{k,X'}\int^{p,X'}_{k',X''}(X_0''-X_0')\big(\partial_{X''}^{1}\partial_{X''[1}\langle E_{1}^a(X') E_{2]}^a (X'')\rangle
+\partial_{X''}^{2}\partial_{X''[1}\langle E_{2}^a(X') E_{2]}^a (X'')\rangle\big).
\end{eqnarray}
By using Eq.~(\ref{eq:rel_DE_DB}), $\mathcal{A}_{b}$ reduces to 
\begin{eqnarray}
\mathcal{A}_{b}=+\int^{p,X}_{k,X'}\int^{p,X'}_{k',X''}\big(\partial^{1}_{X'}\langle B^{a}_{3}(X')E^{a}_{1}(X'')\rangle
+\partial^{2}_{X'}\langle B^{a}_{3}(X')E^{a}_{2}(X'')\rangle\big).
\end{eqnarray} 

From Eq.~(\ref{eq:useful_int}), one finds
\begin{eqnarray}\nonumber
\mathcal{A}_{a}&=&\frac{\bar{N}_c(N_c^2-1)}{4p_0^2}\int^{X_0}_{X'_0}\int^{X'_0}_{X''_0}\int^{X}_{\perp;q,u}\int^{X}_{\perp;l,v}\Omega_{-}(u_{\perp},v_{\perp})\frac{q^x l^y}{ql}
\partial_{X''0}\Big[\big(J_1(qX'_0)J_1(lX''_0)-J_1(qX''_0)J_1(lX'_0)\big)
\\
&&\times\Theta(X'_0)\Theta(X''_0)\Big],
\end{eqnarray}
where we have further used
\begin{eqnarray}
	\int^{X_0}_{X'_0}\equiv \int^{\infty}_{-\infty} d X'_0\Big(1+{\rm sgn}(X_0-X_0')\Big)=2\int^{\infty}_{-\infty} d X'_0\Theta(X_0-X_0').
\end{eqnarray}
Here we have implicitly multiplied the color-field correlators with the unit-step functions for $X_0'$ and $X_0''$, $\Theta(X'_0)$ and $\Theta(X''_0)$, because of setting $X_0'$=$X_0''=0$ as an initial time for the presence of glamsa.
It turns out that $\mathcal{A}_{a}=0$ due to the structure $\langle E_{[2}^a(X') E_{1]}^a (X'')\rangle$ since 
\begin{eqnarray}\nonumber
\int^{X'_0}_{X''_0}\partial_{X''0}J_1(qX'_0)J_1(lX''_0)\Theta(X'_0)\Theta(X''_0)&=&\int^{X'_0}_{X''_0}\partial_{X''0}J_1(qX''_0)J_1(lX'_0)\Theta(X'_0)\Theta(X''_0)
\\
&=&2J_1(qX'_0)J_1(lX'_0)\Theta(X'_0)
.
\end{eqnarray}
On the other hand, it is found
\begin{eqnarray}\nonumber
	\mathcal{A}_{b}
	&=&-\frac{i\bar{N}_c(N_c^2-1)}{4p_0^2}\int^{X_0}_{X'_0}\int^{X'_0}_{X''_0}\bigg[\partial^1_{X'}\int^{X'}_{\perp;q,u}\int^{X''}_{\perp;l,v}\Omega_{-}(u_{\perp},v_{\perp})\frac{l^y}{l}J_0(qX'_0)J_1(lX''_0)
	\\
	&&-\partial^2_{X'}\int^{X'}_{\perp;q,u}\int^{X''}_{\perp;l,v}\Omega_{-}(u_{\perp},v_{\perp})\frac{l^x}{l}J_0(qX'_0)J_1(lX''_0)\bigg]_{X''_{\perp}=X'_{\perp}=X_{\perp}}\Theta(X'_0)\Theta(X''_0).
\end{eqnarray}
To handle the spatial derivative terms, we introduce the trick,
\begin{eqnarray}\nonumber
	\partial_{X'_{\perp}j}\int^{X'}_{\perp;q,u}G(u_{\perp})&=&\int \!\frac{d^2q_{\perp}}{(2\pi)^2}\int \!d^2u_{\perp}(-iq_{\perp j})e^{iq_{\perp}(X'-u)_{\perp}}G(u_{\perp})
	\\
	&=&-\int \!\frac{d^2q_{\perp}}{(2\pi)^2}\int \!d^2u_{\perp}\big(\partial_{u_{\perp}j}e^{iq_{\perp}(X'-u)_{\perp}}\big)G(u_{\perp}),
\end{eqnarray}
for an arbitrary function $G(u_{\perp})$. When assuming $G(u_{\perp})|_{u^i_{\perp}\rightarrow \pm \infty}\rightarrow 0$, we further obtain the useful relation, 
\begin{eqnarray}\label{eq:derivative_trick}
	\partial_{X'_{\perp}j}\int^{X'}_{\perp;q,u}G(u_{\perp})=\int^{X'}_{\perp;q,u}\partial_{u_{\perp}j}G(u_{\perp}).
\end{eqnarray}
Assuming  
\begin{eqnarray}
	\Omega_{\pm}(u_{\perp},v_{\perp})|_{u^i_{\perp}\rightarrow \pm \infty}\rightarrow 0,\quad \Omega_{\pm}(u_{\perp},v_{\perp})|_{v^i_{\perp}\rightarrow \pm \infty}\rightarrow 0,
\end{eqnarray}
which are physical boundary conditions, we accordingly derive 
\begin{eqnarray}
\mathcal{A}_{b}=-\frac{i\bar{N}_c(N_c^2-1)}{4p_0^2}\int^{X_0}_{X'_0}\int^{X'_0}_{X''_0}\int^{X}_{\perp;q,u}\int^{X}_{\perp;l,v}\frac{J_0(qX'_0)J_1(lX''_0)}{l}l^{[y}\partial^{x]}_{u}\Omega_{-}(u_{\perp},v_{\perp})\Theta(X'_0)\Theta(X''_0)
\end{eqnarray}
and similarly
\begin{eqnarray}\nonumber
\mathcal{A}_{c}&=&\frac{\bar{N}_c(N_c^2-1)}{4p_0^2}\int^{X_0}_{X'_0}\int^{X'_0}_{X''_0}(X_0''-X'_0)\int^{X}_{\perp;q,u}\int^{X}_{\perp;l,v}\frac{J_1(qX'_0)J_1(lX''_0)}{ql}\Theta(X'_0)\Theta(X''_0)
\\
&&\times \big[q^{y}l^{x}\partial^{x}_{v}\partial_{vx}-q^{x}l^{y}\partial^{y}_{v}\partial_{vy}
+(q^yl^y-q^xl^x)\partial^{x}_{v}\partial_{vy}
\big]\Omega_{-}(u_{\perp},v_{\perp}).
\end{eqnarray}
Implementing the decomposition in Eq.~(\ref{eq:decomp_qi}) and integrals in Eq.~(\ref{eq:thetaint_1}) to cope with the angular parts of momentum integration, one finds
\begin{eqnarray}\nonumber
\mathcal{A}_{b}&=&\frac{\bar{N}_c(N_c^2-1)}{4p_0^2}\int^{X_0}_{X'_0}\int^{X'_0}_{X''_0}\int\frac{dqdl ql}{(2\pi)^2}\int d^2u_{\perp}d^2v_{\perp}J_0(qX'_0)J_1(lX''_0)\Theta(X'_0)\Theta(X''_0)
\\
&&\times J_0(q|\bar{u}_{\perp}|)J_1(l|\bar{v}_{\perp}|)\frac{\bar{v}_{\perp}^{[y}\partial^{x]}_{u}}{|\bar{v}_{\perp}|}\Omega_{-}(u_{\perp},v_{\perp})
\end{eqnarray}
and 
\begin{eqnarray}\nonumber
\mathcal{A}_{c}&=&-\frac{\bar{N}_c(N_c^2-1)}{4p_0^2}\int^{X_0}_{X'_0}\int^{X'_0}_{X''_0}(X_0''-X'_0)\int\frac{dqdl ql}{(2\pi)^2}\int d^2u_{\perp}d^2v_{\perp}\frac{J_1(qX'_0)J_1(lX''_0)}{|\bar{u}_{\perp}||\bar{v}_{\perp}|}\Theta(X'_0)\Theta(X''_0)
\\
&&\times J_1(q|\bar{u}_{\perp}|)J_1(l|\bar{v}_{\perp}|) \big[\bar{u}_{\perp}^{y}\bar{v}_{\perp}^{x}\partial^{x}_{v}\partial_{vx}-\bar{u}_{\perp}^{x}\bar{v}_{\perp}^{y}\partial^{y}_{v}\partial_{vy}
+(\bar{u}_{\perp}^y\bar{v}_{\perp}^y-\bar{u}_{\perp}^x\bar{v}_{\perp}^x)\partial^{x}_{v}\partial_{vy}
\big]\Omega_{-}(u_{\perp},v_{\perp}),
\end{eqnarray}
where $\bar{u}_{\perp}=X_{\perp}-u_{\perp}$ and $\bar{v}_{\perp}=X_{\perp}-v_{\perp}$. Further applying the orthogonal condition for Bessel functions, 
\begin{eqnarray}\label{eq:Bessel_rel}
	\int^{\infty}_{0}dr rJ_{\nu}(kr)J_{\nu}(sr)=\frac{\delta(k-s)}{s},
\end{eqnarray}
$\mathcal{A}_{b}$ and $\mathcal{A}_{c}$ become
\begin{eqnarray}\label{eq:Ab_final}
\mathcal{A}_{b}=\frac{g^2N_c(N_c^2-1)}{2(2\pi)^2p_0^2}\int^{\perp}_{\bar{u},\bar{v}}\frac{\Theta(X_0-|\bar{u}_{\perp}|)\Theta(|\bar{u}_{\perp}|-|\bar{v}_{\perp}|)}{|\bar{u}_{\perp}||\bar{v}_{\perp}|}
\frac{\bar{v}_{\perp}^{[y}\partial^{x]}_{u}}{|\bar{v}_{\perp}|}\Omega_{-}(u_{\perp},v_{\perp})
\end{eqnarray}
and
\begin{eqnarray}\nonumber\label{eq:Ac_final}
\mathcal{A}_{c}&=&\frac{-g^2N_c(N_c^2-1)}{2(2\pi)^2p_0^2}\int^{\perp}_{\bar{u},\bar{v}}\frac{\Theta(X_0-|\bar{u}_{\perp}|)\Theta(|\bar{u}_{\perp}|-|\bar{v}_{\perp}|)}{|\bar{u}_{\perp}|^2|\bar{v}_{\perp}|^2}(|\bar{v}_{\perp}|-|\bar{u}_{\perp}|)
\\
&&\times  \big[\bar{u}_{\perp}^{y}\bar{v}_{\perp}^{x}\partial^{x}_{v}\partial_{vx}-\bar{u}_{\perp}^{x}\bar{v}_{\perp}^{y}\partial^{y}_{v}\partial_{vy}
+(\bar{u}_{\perp}^y\bar{v}_{\perp}^y-\bar{u}_{\perp}^x\bar{v}_{\perp}^x)\partial^{x}_{v}\partial_{vy}
\big]\Omega_{-}(u_{\perp},v_{\perp}),
\end{eqnarray}
where we have implicitly taken $\Theta(|\bar{u}_{\perp}|)\Theta(|\bar{v}_{\perp}|)=1$ and introduced 
\begin{eqnarray}
\int^{\perp}_{u,v}\equiv\int d^2u_{\perp}d^2v_{\perp}.
\end{eqnarray}
Given an explicit expression of $\Omega_{-}(u_{\perp},v_{\perp})$, Eqs.~(\ref{eq:Ab_final}) and (\ref{eq:Ac_final}) can be evaluated numerically.

\subsection{Spin correlation}\label{sec:spin_corr}
We may now evaluate the spin correlation $\langle\mathcal{P}^{y}_{q}\mathcal{P}^{y}_{\bar{q}}\rangle$ dynamically generated by $\langle\tilde{a}^{sy}(p,X)\tilde{a}^{sy}(p,Y)\rangle$. By symmetry, $\langle\tilde{a}^{sy}(p,X)\tilde{a}^{sy}(p,Y)\rangle=\langle\tilde{a}^{sy}(p,Y)\tilde{a}^{sy}(p,X)\rangle$ because we will in the end integrate over spatial $X$ and $Y$ on the freeze-out hyper-surface with $X_0=Y_0$, the integrand in the multiple integrals involved is also invariant under $(X'\leftrightarrow Y',X''\leftrightarrow Y'')$. The explicit form of $\langle\tilde{a}^{sy}(p,X)\tilde{a}^{sy}(p,Y)\rangle$ can be written as
\begin{eqnarray}\nonumber\label{eq:ayay}
	\langle\tilde{a}^{sy}(p,X)\tilde{a}^{sy}(p,Y)\rangle
	&\approx&\Bigg(\frac{\tilde{C}_2}{2}
	p_0\big(\partial_{p0}f_V(p_0)\big)\Bigg)^2\int^{p,X}_{k,X'}\int^{p,X'}_{k',X''}\int^{p,Y}_{\bar{k},Y'}\int^{p,Y'}_{\bar{k}',Y''}\nonumber
	\\&&\Big[\partial_{X''0}\partial_{Y''0}\left\langle E_{[1}^a(X') E_{3]}^a (X'')E_{[1}^b(Y') E_{3]}^b (Y'')\right\rangle \nonumber
	\\&& -2\partial_{X''0}\partial_{Y'1}\left\langle E_{[1}^a(X') E_{3]}^a (X'')B^{b[2}(Y') E^{b1]} (Y'')\right\rangle
	\nonumber\\&&
	+2(Y_0''-Y_0')\partial_{X''0}\partial^2_{Y''1}\left\langle E_{[1}^a(X') E_{3]}^a (X'')E_1^b(Y') E_3^b (Y'')\right\rangle
	\nonumber\\&&+2(Y_0''-Y_0')\partial_{X''0}\partial_{Y''2}\partial_{Y''1}\left\langle E_{[1}^a(X') E_{3]}^a (X'')E_2^b(Y') E_3^b(Y'')\right\rangle
	\nonumber\\&&
	+\partial_{X'1}\partial_{Y'1}\left\langle B^{a[2}(X')E^{a1]}(X'')B^{b[2}(Y')E^{b1]}(Y'')\right\rangle	\nonumber\\&&
	-2(Y_0''-Y_0')\partial_{X'1}\partial^2_{Y''1}\left\langle B^{a[2}(X')E^{a1]}(X'')E_1^b(Y') E_3^b (Y'')\right\rangle
	\nonumber\\&&-2(Y_0''-Y_0')\partial_{X'1}\partial_{Y''2}\partial_{Y''1}\left\langle B^{a[2}(X')E^{a1]}(X'')E_2^b(Y') E_3^b (Y'')\right\rangle
	\nonumber\\&&
	+(X_0''-X_0')(Y_0''-Y_0')\Big(\partial^2_{X''1}\partial^2_{Y''1}\left\langle E_1^a(X') E_3^a (X'')E_{1}^b(Y') E_{3}^b (Y'')\right\rangle\nonumber\\
	&&+2\partial^2_{X''1}\partial_{Y''2}\partial_{Y''1}\left\langle E_1^a(X') E_3^a (X'')E_2^b(Y') E_3^b (Y'')\right\rangle\nonumber\\
	&&+\partial_{X''2}\partial_{X''1}\partial_{Y''2}\partial_{Y''1}\left\langle E_2^a(X') E_3^a (X'')E_2^b(Y') E_3^b (Y'')\right\rangle\Big)\Big].
\end{eqnarray}
Nonetheless, one has to derive a more concise form for the multi-dimensional integral with the input of color-field correlators. Such an analytic expression for each term above could be derived in light of the procedure shown in appendix~\ref{app:spin_corr}, where we take $\langle E_{1}^a(X') E_{3}^a (X'')E_{1}^b(Y') E_{3}^b (Y'')\rangle$ as an example.  
It turns out that all the terms associated with $\partial_{X0''} E_{[1}^a(X') E_{3]}^a (X'')$ in Eq.~(\ref{eq:ayay}) vanish since such terms involve the integral
\begin{eqnarray}
	\int^{X'_0}_{X_0''}\partial_{X0''}J_1(qX_0'')J_0(lX'_0)\Theta(X_0')\Theta(X_0'')-
	\int^{X'_0}_{X_0''}\partial_{X0''}J_1(qX_0')J_0(lX''_0)\Theta(X_0')\Theta(X_0'')=0,
\end{eqnarray}
which is similar to the reason causing $\mathcal{A}_{a}=0$.
One may refer to appendix~\ref{app:vanishing_terms} for an explicit calculation showing that e.g. the contribution from the second line of Eq.~(\ref{eq:ayay}) vanishes.
Eventually, Eq.~(\ref{eq:ayay}) reduces to
\begin{eqnarray}\label{eq:ayays}
	\langle\tilde{a}^{sy}(p,X)\tilde{a}^{sy}(p,Y)\rangle
	&\approx&\Bigg(\frac{\tilde{C}_2}{2}
	p_0\big(\partial_{p0}f_V(p_0)\big)\Bigg)^2\big(\mathcal{I}_1+\mathcal{I}_2+\mathcal{I}_3\big),
\end{eqnarray} 
where
\begin{eqnarray}
\mathcal{I}_1=\int^{p,X}_{k,X'}\int^{p,X'}_{k',X''}\int^{p,Y}_{\bar{k},Y'}\int^{p,Y'}_{\bar{k}',Y''}\partial_{X'1}\partial_{Y'1}\left\langle B^{a[2}(X')E^{a1]}(X'')B^{b[2}(Y')E^{b1]}(Y'')\right\rangle,
\end{eqnarray}
\begin{eqnarray}\nonumber
\mathcal{I}_{2}&=& -2\int^{p,X}_{k,X'}\int^{p,X'}_{k',X''}\int^{p,Y}_{\bar{k},Y'}\int^{p,Y'}_{\bar{k}',Y''}(Y_0''-Y_0')\Big(\partial_{X'1}\partial^2_{Y''1}\left\langle B^{a[2}(X')E^{a1]}(X'')E_1^b(Y') E_3^b (Y'')\right\rangle
\\
&&+\partial_{X'1}\partial_{Y''2}\partial_{Y''1}\left\langle B^{a[2}(X')E^{a1]}(X'')E_2^b(Y') E_3^b (Y'')\right\rangle\Big),
\end{eqnarray}
and
\begin{eqnarray}
\mathcal{I}_{3}&=& \int^{p,X}_{k,X'}\int^{p,X'}_{k',X''}\int^{p,Y}_{\bar{k},Y'}\int^{p,Y'}_{\bar{k}',Y''}(X_0''-X_0')(Y_0''-Y_0')\Big(\partial^2_{X''1}\partial^2_{Y''1}\left\langle E_1^a(X') E_3^a (X'')E_{1}^b(Y') E_{3}^b (Y'')\right\rangle\nonumber\\
&&+2\partial^2_{X''1}\partial_{Y''2}\partial_{Y''1}\left\langle E_1^a(X') E_3^a (X'')E_2^b(Y') E_3^b (Y'')\right\rangle\nonumber\\
&&+\partial_{X''2}\partial_{X''1}\partial_{Y''2}\partial_{Y''1}\left\langle E_2^a(X') E_3^a (X'')E_2^b(Y') E_3^b(Y'')\right\rangle\Big).
\end{eqnarray}
It is found that the non-vanishing spin correlation stems from the transverse spatial derivatives upon color-field correlators. 

We will compute $\mathcal{I}_1$ as an example, while the other terms in Eq.~(\ref{eq:ayays}) can be derived in a similar manner. By symmetry, the integrand of $\mathcal{I}_1$ can be written as
\begin{eqnarray}\nonumber
&&\partial_{X'1}\partial_{Y'1}\left\langle B^{a[2}(X')E^{a1]}(X'')B^{b[2}(Y')E^{b1]}(Y'')\right\rangle
\\\nonumber
&&= \partial_{X'1}\partial_{Y'1}\big(\left\langle B^{a2}(X')B^{b2}(Y')\right\rangle\left\langle E^{a1}(X'')E^{b1}(Y'')\right\rangle
+\left\langle B^{a1}(X')B^{b1}(Y')\right\rangle\left\langle E^{a2}(X'')E^{b2}(Y'')\right\rangle
\\
&&\quad-2\left\langle B^{a2}(X')B^{b1}(Y')\right\rangle\left\langle E^{a1}(X'')E^{b2}(Y'')\right\rangle\big),
\end{eqnarray} 
where we have applied the Wick-theorem like decomposition to decompose a four-field correlator in terms of the products of two-field correlators, e.g.  
\begin{eqnarray}\nonumber
&&\left\langle B^{a2}(X')E^{a1}(X'')B^{b2}(Y')E^{b1}(Y'')\right\rangle
\\\nonumber
&&=\left\langle B^{a2}(X')E^{a1}(X'')\right\rangle\left\langle B^{b2}(Y')E^{b1}(Y'')\right\rangle
+\left\langle B^{a2}(X')B^{b2}(Y')\right\rangle\left\langle E^{a1}(X'')E^{b1}(Y'')\right\rangle
\\
&&\quad+\left\langle B^{a2}(X')E^{b1}(Y'')\right\rangle\left\langle E^{a1}(X'')B^{b2}(Y')\right\rangle.
\end{eqnarray}
We then make the further decomposition, $\mathcal{I}_{1}=\mathcal{I}_{1a}+\mathcal{I}_{1b}-2\mathcal{I}_{1c}$, where
\begin{eqnarray}
\mathcal{I}_{1a}\equiv \int^{p,X}_{k,X'}\int^{p,X'}_{k',X''}\int^{p,Y}_{\bar{k},Y'}\int^{p,Y'}_{\bar{k}',Y''}\partial_{X'1}\partial_{Y'1}\left\langle B^{a2}(X')B^{b2}(Y')\right\rangle\left\langle E^{a1}(X'')E^{b1}(Y'')\right\rangle,
\end{eqnarray}
\begin{eqnarray}
	\mathcal{I}_{1b}\equiv \int^{p,X}_{k,X'}\int^{p,X'}_{k',X''}\int^{p,Y}_{\bar{k},Y'}\int^{p,Y'}_{\bar{k}',Y''}\partial_{X'1}\partial_{Y'1}\left\langle B^{a1}(X')B^{b1}(Y')\right\rangle\left\langle E^{a2}(X'')E^{b2}(Y'')\right\rangle,
\end{eqnarray}
and
\begin{eqnarray}
	\mathcal{I}_{1c}\equiv \int^{p,X}_{k,X'}\int^{p,X'}_{k',X''}\int^{p,Y}_{\bar{k},Y'}\int^{p,Y'}_{\bar{k}',Y''}\partial_{X'1}\partial_{Y'1}\left\langle B^{a2}(X')B^{b1}(Y')\right\rangle\left\langle E^{a1}(X'')E^{b2}(Y'')\right\rangle.
\end{eqnarray}

To evaluate $\mathcal{I}_1$, we shall apply the same tricks in Eq.~(\ref{eq:useful_int}) and Eq.~(\ref{eq:derivative_trick}), which yield
\begin{eqnarray}\nonumber
	\mathcal{I}_{1a}(p,X,Y)&\approx& \frac{g^4N_c^2(N_c^2-1)}{64p_0^4}
	\int^{X}_{\perp;q,u}\int^{Y}_{\perp;l,v}\int^{X}_{\perp;q',u'}\int^{Y}_{\perp;l',v'}\left(\frac{q^xl^x}{q l}\right)\left(\frac{q'^yl'^y}{q' l'}\right)
	\\
	&&\times\partial_{ux}\partial_{vx}\Omega_{+}(u_{\perp},v_{\perp})\Omega_{-}(u'_{\perp},v'_{\perp})\mathcal{Y}_{1a}(X_0,Y_0,q,l,q',l'),
\end{eqnarray}
\begin{eqnarray}\nonumber
	\mathcal{I}_{1b}(p,X,Y)&\approx& \frac{g^4N_c^2(N_c^2-1)}{64p_0^4}
	\int^{X}_{\perp;q,u}\int^{Y}_{\perp;l,v}\int^{X}_{\perp;q',u'}\int^{Y}_{\perp;l',v'}\left(\frac{q^yl^y}{q l}\right)\left(\frac{q'^xl'^x}{q' l'}\right)
	\\
	&&\times\partial_{ux}\partial_{vx}\Omega_{+}(u_{\perp},v_{\perp})\Omega_{-}(u'_{\perp},v'_{\perp})\mathcal{Y}_{1a}(X_0,Y_0,q,l,q',l'),
\end{eqnarray}
and
\begin{eqnarray}\nonumber
	\mathcal{I}_{1c}(p,X,Y)&\approx& \frac{g^4N_c^2(N_c^2-1)}{64p_0^4}
	\int^{X}_{\perp;q,u}\int^{Y}_{\perp;l,v}\int^{X}_{\perp;q',u'}\int^{Y}_{\perp;l',v'}\left(\frac{q^xl^y}{q l}\right)\left(\frac{q'^yl'^x}{q' l'}\right)
	\\
	&&\times\partial_{ux}\partial_{vx}\Omega_{+}(u_{\perp},v_{\perp})\Omega_{-}(u'_{\perp},v'_{\perp})\mathcal{Y}_{1a}(X_0,Y_0,q,l,q',l'),
\end{eqnarray}
where we introduced several shorthand notations,
\begin{eqnarray}
	\int^{X'}_{\perp;q,u}\equiv\int \!\frac{d^2q_{\perp}}{(2\pi)^2}\int \!d^2u_{\perp}e^{iq_{\perp}(X'-u)_{\perp}},
\end{eqnarray}
\begin{eqnarray}\nonumber
	&&\mathcal{Y}_{1a}(X_0,Y_0,q,l,q',l')
	\\
	&&\equiv\int^{X_0}_{X'_0}\int^{Y_0}_{Y'_0}\int^{X'_0}_{X''_0}\int^{Y'_0}_{Y''_0}
	J_1(qX'_0)J_1(lY'_0)\Theta(X'_0)\Theta(Y'_0)
	J_1(q'X''_0)J_1(l'Y''_0)\Theta(X''_0)\Theta(Y''_0),
\end{eqnarray}
and 
\begin{eqnarray}
	s_{\perp}=X_{\perp}-u_{\perp},\quad 
	s'_{\perp}=X_{\perp}-u'_{\perp} \quad
	t_{\perp}=Y_{\perp}-v_{\perp} \quad
	t'_{\perp}=Y_{\perp}-v'_{\perp}.
\end{eqnarray}
One hence obtains
\begin{eqnarray}\nonumber
	\mathcal{I}_{1}
	&\approx&\frac{g^4N_c^2(N_c^2-1)}{64p_0^4}
	\int^{X}_{\perp;q,u}\int^{Y}_{\perp;l,v}\int^{X}_{\perp;q',u'}\int^{Y}_{\perp;l',v'}\partial_{ux}\partial_{vx}\Omega_{+}(u_{\perp},v_{\perp})\Omega_{-}(u'_{\perp},v'_{\perp})\mathcal{Y}_{1a}(X_0,Y_0,q,l,q',l')
	\\
	&&\quad\times\frac{1}{qlq'l'}\big[q^xl^xq^{\prime y}l^{\prime y}+q^yl^yq^{\prime x}l^{\prime x}-2q^xl^yq^{\prime y}l^{\prime x}\big].
\end{eqnarray}
Utilizing Eq.~(\ref{eq:decomp_qi}) and Eq.~(\ref{eq:thetaint_1}) again to integrate over the angular parts of momentum integrals, $\mathcal{I}_{1}$ becomes
\begin{eqnarray}\nonumber
	\mathcal{I}_{1}&\approx& \frac{g^4N_c^2(N_c^2-1)}{4p_0^4}
\int \!dq\int \!dl\int \!\frac{dq'}{(2\pi)^2}\int \!\frac{dl'}{(2\pi)^2}\int^{\perp}_{u,v,u',v'}qlq'l'
	\\\nonumber
	&&\times \big[\hat{s}_{\perp}^{x}\hat{t}_{\perp}^{x}\hat{s}_{\perp}^{\prime y}\hat{t}_{\perp}^{\prime y}+\hat{s}_{\perp}^{y}\hat{t}_{\perp}^{y}\hat{s}_{\perp}^{\prime x}\hat{t}_{\perp}^{\prime x}-2\hat{s}_{\perp}^{x}\hat{t}_{\perp}^{y}\hat{s}_{\perp}^{\prime y}\hat{t}_{\perp}^{\prime x}\big]J_1(q|s_{\perp}|)J_1(l|t_{\perp}|)J_1(q'|s'_{\perp}|)J_1(l'|t'_{\perp}|)
	\\
	&&\times \partial_{ux}\partial_{vx}\Omega_{+}(u_{\perp},v_{\perp})\Omega_{-}(u'_{\perp},v'_{\perp})J_1(qX'_0)J_1(lY'_0)
	J_1(q'X''_0)J_1(l'Y''_0),
\end{eqnarray}
where $\hat{s}^i_{\perp}=s^i_{\perp}/|s_{\perp}|$ and
\begin{eqnarray}
	\int^{\perp}_{u,v,u'v'}\equiv \int \!d^2u_{\perp}\!\!\int \!d^2v_{\perp}
	\int \!d^2u'_{\perp}\!\!\int \!d^2v'_{\perp}.
\end{eqnarray}
Further using Eq.~(\ref{eq:Bessel_rel}), we arrive at
\begin{eqnarray}\nonumber\label{eq:I1_final}
	\mathcal{I}_{1}&\approx& \frac{g^4N_c^2(N_c^2-1)}{4(2\pi)^4p_0^4}
	\int^{\perp}_{s,t,s',t'}\frac{\Theta(X_0-|s_{\perp}|)\Theta(Y_0-|t_{\perp}|)\Theta(|s_{\perp}|-|s'_{\perp}|)\Theta(|t_{\perp}|-|t'_{\perp}|)}{|s_{\perp}||t_{\perp}||s'_{\perp}||t'_{\perp}|}
	\\
	&&\times \big[\hat{s}_{\perp}^{x}\hat{t}_{\perp}^{x}\hat{s}_{\perp}^{\prime y}\hat{t}_{\perp}^{\prime y}+\hat{s}_{\perp}^{y}\hat{t}_{\perp}^{y}\hat{s}_{\perp}^{\prime x}\hat{t}_{\perp}^{\prime x}-2\hat{s}_{\perp}^{x}\hat{t}_{\perp}^{y}\hat{s}_{\perp}^{\prime y}\hat{t}_{\perp}^{\prime x}\big]
    \partial_{ux}\partial_{vx}\Omega_{+}(u_{\perp},v_{\perp})\Omega_{-}(u'_{\perp},v'_{\perp}),
\end{eqnarray}
which could be evaluated numerically with given $\Omega_{\pm}$. Following the same procedure, one can also derive similar expressions for the remaining terms in Eq.~(\ref{eq:ayays}). As shown in appendix.~\ref{app:remaining_terms}, it is found
\begin{eqnarray}\nonumber\label{eq:I2_final}
\mathcal{I}_{2}
&\approx& \frac{g^4N_c^2(N_c^2-1)}{2(2\pi)^4p_0^4}
\int^{\perp}_{s,t,s',t'}\frac{\Theta(X_0-|s_{\perp}|)\Theta(Y_0-|t_{\perp}|)\Theta(|s_{\perp}|-|s'_{\perp}|)\Theta(|t'_{\perp}|-|t_{\perp}|)}{|s_{\perp}||t_{\perp}||s'_{\perp}||t'_{\perp}|}
\\
&&\times\big(|t_{\perp}|-|t'_{\perp}|\big)\hat{s}_{\perp}^{[x} \hat{s}'^{ y]}_{\perp}\big(\hat{t}'^y_{\perp}\partial_{vx}-\hat{t}'^x_{\perp}\partial_{vy}\big)\partial_{ux}\partial_{vx}
\Omega_{-}(v'_{\perp},u'_{\perp})\Omega_{+}(u_{\perp},v_{\perp})
\end{eqnarray}
and
\begin{eqnarray}\nonumber\label{eq:I3_final}
\mathcal{I}_{3}
	&\approx& -\frac{g^4N_c^2(N_c^2-1)}{4(2\pi)^4p_0^4}
\int^{\perp}_{s,t,s',t'}\frac{\Theta(X_0-|s_{\perp}|)\Theta(Y_0-|t_{\perp}|)\Theta(|s_{\perp}|-|s'_{\perp}|)\Theta(|t_{\perp}|-|t'_{\perp}|)}{|s_{\perp}||t_{\perp}||s'_{\perp}||t'_{\perp}|}
\\\nonumber
&&\times\big(|s'_{\perp}|-|s_{\perp}|\big)\big(|t'_{\perp}|-|t_{\perp}|\big)
\big[\hat{s}^y_{\perp}\hat{t}^y_{\perp}\partial^2_{u'x}\partial^2_{v'x}-2\hat{s}^y_{\perp}\hat{t}^x_{\perp}\partial^2_{u'x}\partial_{v'y}\partial_{v'x}
\\
&&+\hat{s}^x_{\perp}\hat{t}^x_{\perp}\partial_{u'x}\partial_{u'y}\partial_{v'x}\partial_{v'y}\big]
\Omega_{-}(u_{\perp},v_{\perp})\Omega_{+}(u'_{\perp},v'_{\perp}).
\end{eqnarray}
The major difference is that the terms $(Y_0''-Y_0')$ and $(X_0''-X_0')$ give rise to $\big(|t'_{\perp}|-|t_{\perp}|\big)$ and $\big(|s'_{\perp}|-|s_{\perp}|\big)$ in the remaing integrands after integrating over $Y'_0$, $Y''_0$, $X'_0$, and $X''_0$.

\section{Analysis with the GBW distribution}\label{sec:GBW}

Following Ref.~\cite{Guerrero-Rodriguez:2021ask}, we adopt the GBW distribution such that $h_{1,2}=0$ and 
\begin{align}
	\Omega_{\pm}(u_{\perp},v_{\perp})=\Omega(u_{\perp},v_{\perp})=\frac{Q_s^4}{g^4N_c^2}\left(\frac{1-e^{-Q_s^2|u_{\perp}-v_{\perp}|^2/4}}{Q_s^2|u_{\perp}-v_{\perp}|^2/4}\right)^2=\frac{Q_s^4}{g^4N_c^2}\left(\frac{1-e^{-Q_s^2|s_{\perp}-t_{\perp}-r_{\perp}|^2/4}}{Q_s^2|s_{\perp}-t_{\perp}-r_{\perp}|^2/4}\right)^2,
\end{align}
where $r_{\perp}\equiv X_{\perp}-Y_{\perp}$ and $Q_s$ denotes the saturation momentum. One finds $\mathcal{A}_{b}=0$ and $\mathcal{A}_{c}=0$ after conducting the numerical calculations with the GBW distribution and hence $\langle\tilde{a}^{sz}(p,X)\rangle$=0. It turns out that the spin polarization from glasma vanishes in all directions. We may analytically show $\mathcal{A}_{b}=0$ as an example in the following. From the GBW distribution, it is found
\begin{eqnarray}
	\frac{\bar{v}^{[y}\partial^{x]}_{u}}{|\bar{v}_{\perp}|}\Omega(u_{\perp},v_{\perp})=\mathcal{F}(|\bar{u}_{\perp}|,|\bar{v}_{\perp}|,\cos(\theta_{\bar{v}}-\theta_{\bar{u}}))\sin(\theta_{\bar{v}}-\theta_{\bar{u}})
\end{eqnarray}
where %$|\bar{u}_{\perp}-\bar{v}_{\perp}|^2=|\bar{u}_{\perp}|^2+|\bar{v}_{\perp}^2|-2|\bar{u}_{\perp}||\bar{v}_{\perp}|\cos(\theta_{\bar{v}}-\theta_{\bar{u}})$ and 
$\theta_{\bar{u}}$ and $\theta_{\bar{v}}$ are the polar angles of $\bar{u}_{\perp}$ and $\bar{v}_{\perp}$ and the explicit form of $\mathcal{F}$ is unimportant here. By making the change of coordinates, $\bar{\theta}_{\bar{u},\bar{v}}=\theta_{\bar{u}}+\theta_{\bar{v}}$ and $\Theta_{\bar{u},\bar{v}}=\theta_{\bar{v}}-\theta_{\bar{u}}$, for arbitrary integrand $G(\Theta_{\bar{u},\bar{v}})$ only depending on $\Theta_{\bar{u},\bar{v}}$, it can be shown
\begin{eqnarray}\nonumber
&&\int^{2\pi}_0 d\theta_{\bar{u}}\int^{2\pi}_0 d\theta_{\bar{v}}G(\Theta_{\bar{u},\bar{v}})
\\
&&=2\pi\int^{2\pi}_{-2\pi}d\Theta_{\bar{u},\bar{v}}G(\Theta_{\bar{u},\bar{v}})-\int^{2\pi}_{0}d\Theta_{\bar{u},\bar{v}}\Theta_{\bar{u},\bar{v}}\big(G(\Theta_{\bar{u},\bar{v}})+G(-\Theta_{\bar{u},\bar{v}})\big).
\end{eqnarray} 
It is then clear to see
\begin{eqnarray}
\int^{2\pi}_0 d\theta_{\bar{u}}\int^{2\pi}_0 d\theta_{\bar{v}}\mathcal{F}(|\bar{u}_{\perp}|,|\bar{v}_{\perp}|,\cos\Theta_{\bar{u},\bar{v}})\sin\Theta_{\bar{u},\bar{v}}=0,
\end{eqnarray}
since $\mathcal{F}(|\bar{u}_{\perp}|,|\bar{v}_{\perp}|,\cos\Theta_{\bar{u},\bar{v}})\sin\Theta_{\bar{u},\bar{v}}$ is an odd function under $\Theta_{\bar{u},\bar{v}}\rightarrow -\Theta_{\bar{u},\bar{v}}$.
 
For spin correlation, it turns out that $\mathcal{I}=\mathcal{I}_{1}+\mathcal{I}_{2}+\mathcal{I}_{3}$ only depends on $p_0$, $X_0=Y_0$, and $r_{\perp}$, which is easier to be computed in the polar coordinate.  
Accordingly, $\mathcal{I}$ can be factorized as 
\begin{eqnarray}\label{eq:I_estimate}
	\mathcal{I}(p_0,Q_sX_0,Q_s|r_{\perp}|,\theta_{r})=\frac{g^4N_c^2(N_c^2-1)}{4(2\pi)^4p_0^4}\frac{Q_s^8}{g^8N_c^4}\frac{\hat{\mathcal{I}}(Q_sX_0,Q_s|r_{\perp}|,\theta_{r})}{Q_s^2},
\end{eqnarray}
where $\theta_{r}\equiv \cos^{-1}(r^x_{\perp}/|r_{\perp}|)$ and $\hat{\mathcal{I}}(Q_sX_0,Q_s|r_{\perp}|,\theta_r)$ is the eight-dimensional integral as an dimensionless quantity to be evaluated.
Here the first prefactor associated with $1/p_0^4$ comes from the simplified integral form in Eqs.~(\ref{eq:I1_final}), (\ref{eq:I2_final}), and (\ref{eq:I3_final}), while $Q_s^8/(g^8N_c^4)$ stems from the square of GBW distribution and the extra $1/Q_s^2$ is introduced to make $\hat{\mathcal{I}}(Q_sX_0,Q_s|r_{\perp}|,\theta_r)$ dimensionless.

In addition, we will consider isochronous freeze-out in proper time $\tau=\text{const}$, where $\Sigma_x^{\mu}=(\tau\cosh\eta,x^1,x^2,\tau\sinh\eta)$ \cite{Romatschke:2009im}. In our case, we focus on the small-rapidity region such that $-\eta_{\rm m}\leq\eta\leq \eta_{\rm m}$ with $\eta_{\rm m}\ll 1$. We will also approximate that the space-time rapidity is equal to the
momentum rapidity. Then the normal vector $d\Sigma^{\mu}$ gives
\begin{eqnarray}
	d\Sigma_X\cdot p\approx X_0\sqrt{m^2+|p^2_{\perp}|}d^2X_{\perp}d\eta,
\end{eqnarray}
which yields
\begin{equation}\label{eq:spin_corr}
	\langle\mathcal{P}_q^{i}({\bf p})\mathcal{P}_{\bar{q}}^{i}({\bf p})\rangle \approx \frac{\int d^2X_{\perp}\int d^2Y_{\perp}\langle\tilde{a}^{i}(\bm p,X)\tilde{a}^{i}(\bm p,Y)\rangle}{4m^2\Big(\int d^2X_{\perp}f_{V}(p_0)\Big)^2}\Bigg|_{p_0=\epsilon_{\bm p}}.
\end{equation}
For computational convenience, one may further make the change of coordinates,
\begin{eqnarray}
	\int d^2X_{\perp}\int d^2Y_{\perp}=\int d^2r_{\perp}\int d^2R_{\perp},\quad R_{\perp}\equiv (X_{\perp}+Y_{\perp})/2.
\end{eqnarray} 
Note that $\langle\mathcal{P}_q^{i}({\bf p})\mathcal{P}_{\bar{q}}^{i}({\bf p})\rangle$ depends on $X_0$, which can be chosen as the freeze-out time after which the correlation no longer varies, while the evolution of the numerator and the denominator in Eq.~(\ref{eq:spin_corr}) are different. The subtlety of choosing the freeze-out time will be elaborated below. 

It is important to note that $\langle\tilde{a}^{i}(\bm p,X)\tilde{a}^{i}(\bm p,Y)\rangle$ will stop evolving much earlier than the freeze-out time close to chemical equilibrium in the QGP phase. Furthermore, we have neglected the time dependence of $f_{V}(p_0)$ as the quark distribution function in early times, which should eventually reach thermal equilibrium in the QGP phase. For practical purposes, we should consider different freeze-out hypersurfaces in the numerator and denominator of Eq.~(\ref{eq:spin_corr}), where the freeze-out time in the numerator is chosen to be at thermalization time $X_0=X^{\rm th}_0$ that roughly characterizes the end of the glasma phase, at which the spin no longer evolves, while the one in the numerator is at chemical equilibrium $X_0=X^{\rm eq}_0$. Consequently, from Eq.~(\ref{eq:spin_corr}), the out-of-plane spin correlation may be approximated as
\begin{equation}\label{eq:spin_corr_2}
	\langle\mathcal{P}_q^{y}({\bf p})\mathcal{P}_{\bar{q}}^{y}({\bf p})\rangle \approx \frac{\tilde{C}_2^2p_0^2(\partial_{p0}f_{V})^2\int^{\rm gl} d^2r_{\perp}\int^{\rm gl} d^2R_{\perp}\mathcal{I}(p_0,Q_sX_0,Q_s|r_{\perp}|,\theta_r)}{16m^2{\rm A}_{\rm T}^2f_{\rm eq}(p_0)^2}\Bigg|_{p_0=\epsilon_{\bm p}},
\end{equation} 
where $\int^{\rm gl} d^2r_{\perp}\int^{\rm gl} d^2R_{\perp}$ corresponds to the integrals over the transverse plane of glasma (around the transverse size of collided nuclei) and $\rm A_{\rm T}$ denotes the transverse area of the QGP (around chemical freeze-out). Here $f_{\rm eq}(p_0)=1/(e^{p_0/T}+1)$ corresponds to the thermal distribution function with $T$ being the freeze-out temperature and $m$ may be approximated as the constituent quark mass. We may further make an order-of-magnitude estimation of the spin correlation based on Eq.~(\ref{eq:I_estimate}). Although the exact form of $\hat{\mathcal{I}}(Q_sX_0,Q_s|r_{\perp}|,\theta_r)$ can only be obtained from sophisticated multi-dimensional integral, it is physically expected that the dominant contribution should be around $|r_{\perp}|\sim 0$ with short-range correlation as will be also verified numerically. We accordingly estimate  
\begin{eqnarray}\label{eq:Gaussian_approx}
	\int^{\rm gl} d^2r_{\perp}\mathcal{I}(p_0,Q_sX_0,Q_s|r_{\perp}|,\theta_r)\sim \pi Q_s^{-2}\mathcal{I}(p_0,Q_sX_0,0,0)
\end{eqnarray}
by postulating $\mathcal{I}(p_0,Q_sX_0,Q_s|r_{\perp}|,\theta_r)\sim \mathcal{I}(p_0,Q_sX_0,0,0)e^{-|r_{\perp}|^2Q_s^2}$ as a Gaussian form with the correlation length of $\mathcal{O}(Q_s^{-1})$ and without angular dependence. Taking $\Lambda\sim Q_s\gg m\gg |\bm p|$ such that $\partial_{p0}f_{V}\approx -1/(4Q_s)$ and $\int^{\rm gl} d^2R_{\perp}\approx\rm A_{\rm N}$ with $\rm A_{\rm N}$ being the transverse area of nuclei, from Eq.~(\ref{eq:I_estimate}), Eq.~(\ref{eq:spin_corr_2}) can be further approximated as
\begin{equation}\label{eq:spin_corr_final}
	\langle\mathcal{P}_q^{y}({\bf p})\mathcal{P}_{\bar{q}}^{y}({\bf p})\rangle \sim \frac{(N_c^2-1)(e^{m/T}+1)^2}{N_c^4(16\pi)^4}\frac{\pi Q_s^2\rm A_{\rm N}}{m^4{\rm A}_{\rm T}^2}\hat{\mathcal{I}}(Q_sX^{\rm th}_0,0,0)
\end{equation} 
at small momentum. Despite the actual value of $\hat{\mathcal{I}}(Q_sX^{\rm th}_0,0,0)$, the correlation could be enhanced by the factor $Q_s^2/m^2$ at higher collision energy. Conversely, this effect will be suppressed in low-energy collisions, where the glasma stage does not exist.  

\begin{figure}[t]
	\begin{minipage}{7cm}
		\begin{center}
		{\includegraphics[width=1.1\hsize,height=6cm,clip]{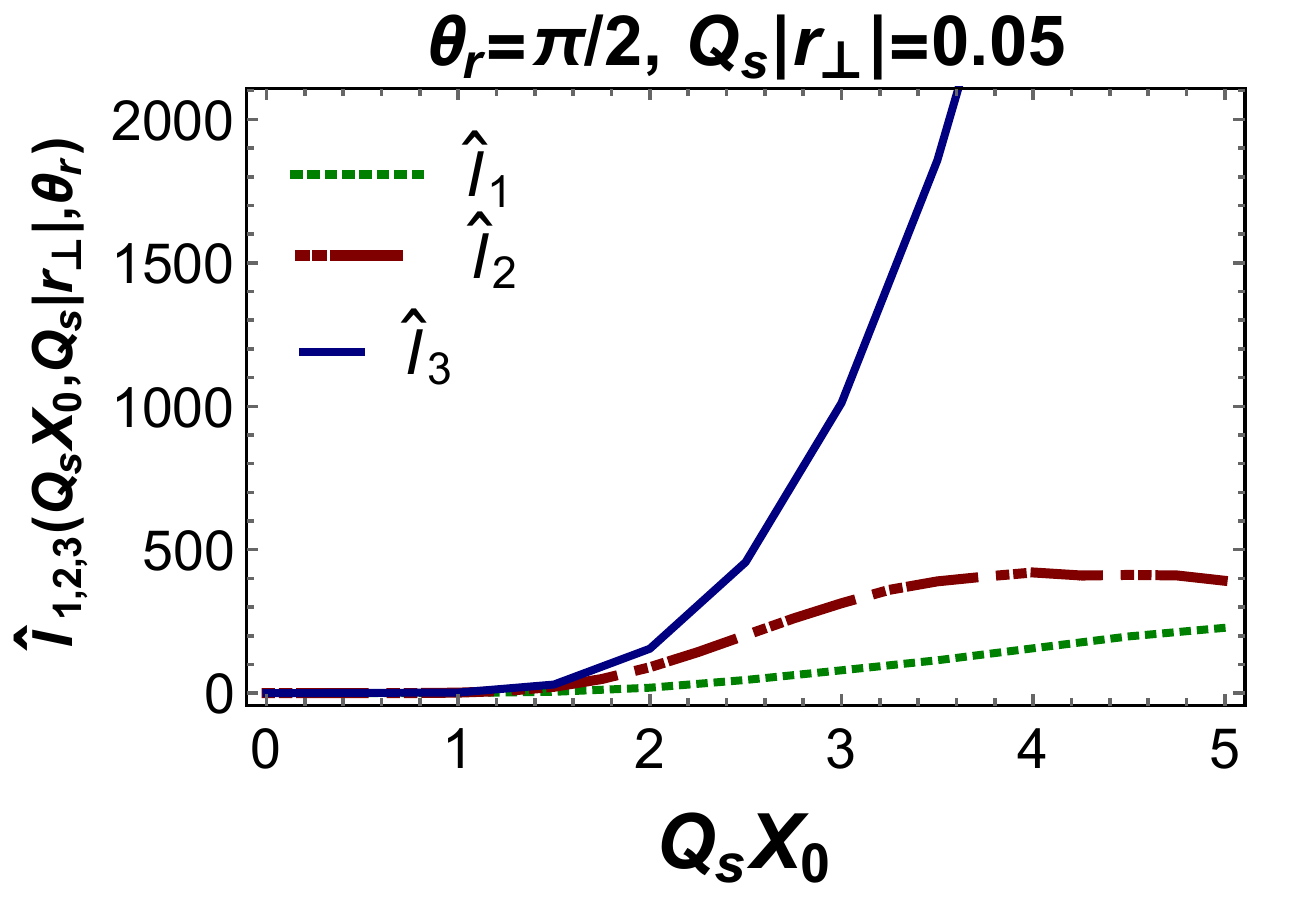}}

			\caption{Numerical results for $\hat{\mathcal{I}}_{1,2,3}(Q_sX_0)$ at $Q_s|r_\perp| = 0.05$ and $\theta_r=\pi/2$. The same behavior is found for other parameter values. 
	}
	\label{figI123}
		\end{center}
	\end{minipage}
	\hspace {1cm}
	\begin{minipage}{7cm}
		\begin{center}
           {\includegraphics[width=1.1\hsize,height=6cm,clip]{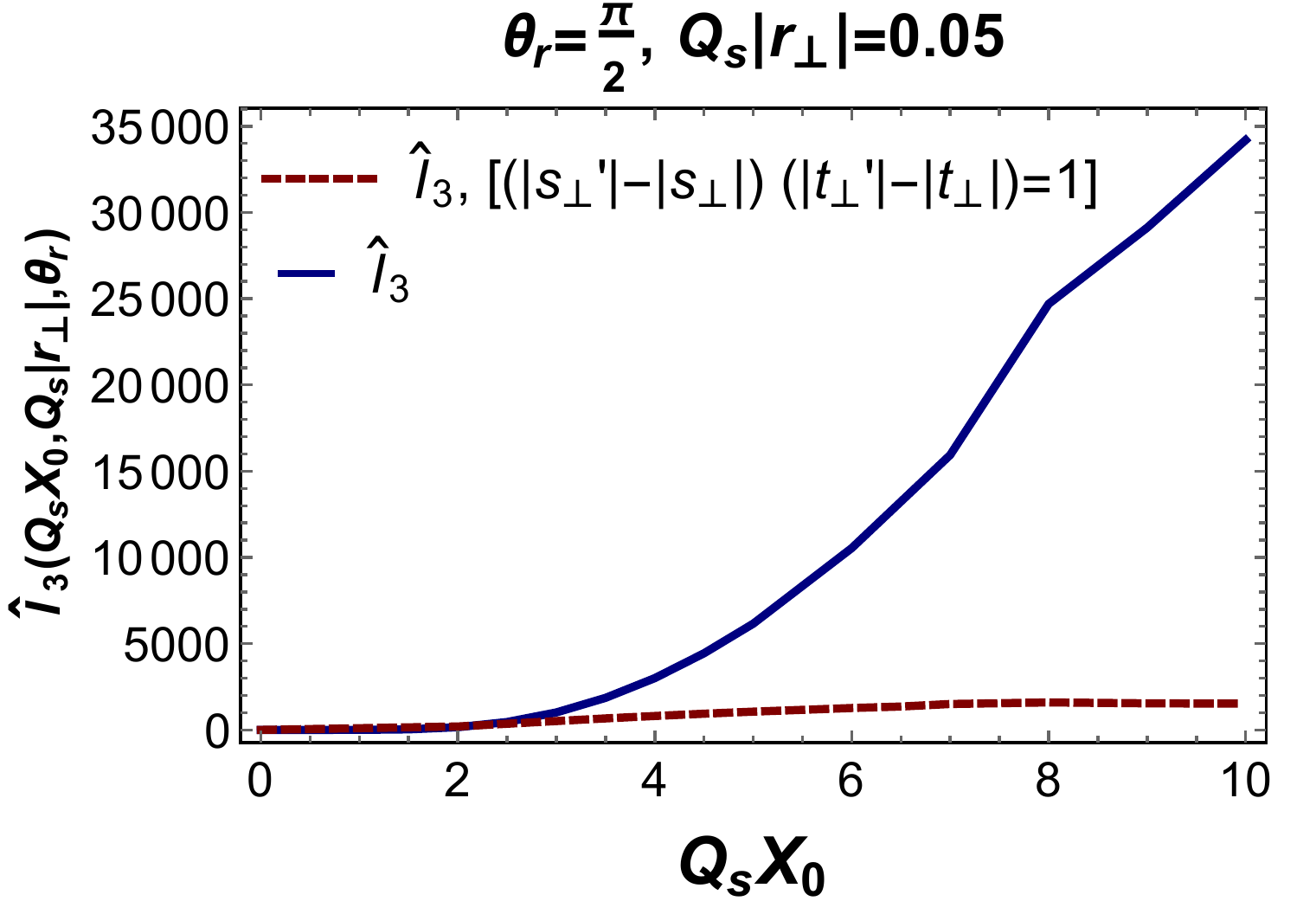}}

			\caption{The origin of the rapid growth for $\hat{I}_3$. The solid line shows the full result; the dashed line shows the result without the secular growth factors.}
			\label{figI3growth}
		\end{center}
	\end{minipage}
\end{figure}

\begin{figure}[t]
	\begin{minipage}{7cm}
		\begin{center}
            {\includegraphics[width=1.1\hsize,height=6cm,clip]{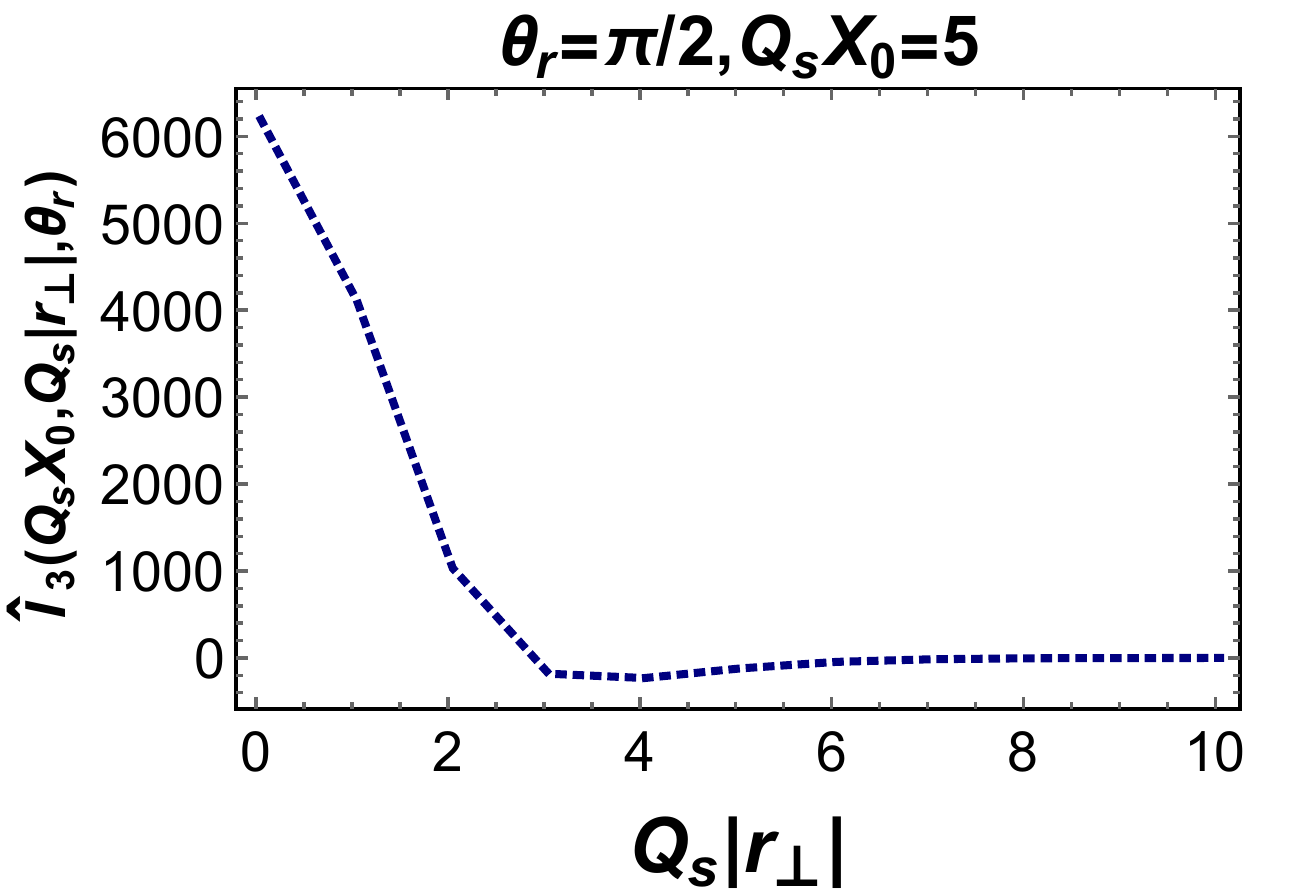}}
			\caption{The $|r_{\perp}|$ dependence of $\hat{I}_3$.}\label{figI3r}
		\end{center}
	\end{minipage}
	\hspace {1cm}
	\begin{minipage}{7cm}
		\begin{center}
   {\includegraphics[width=1.1\hsize,height=6cm,clip]{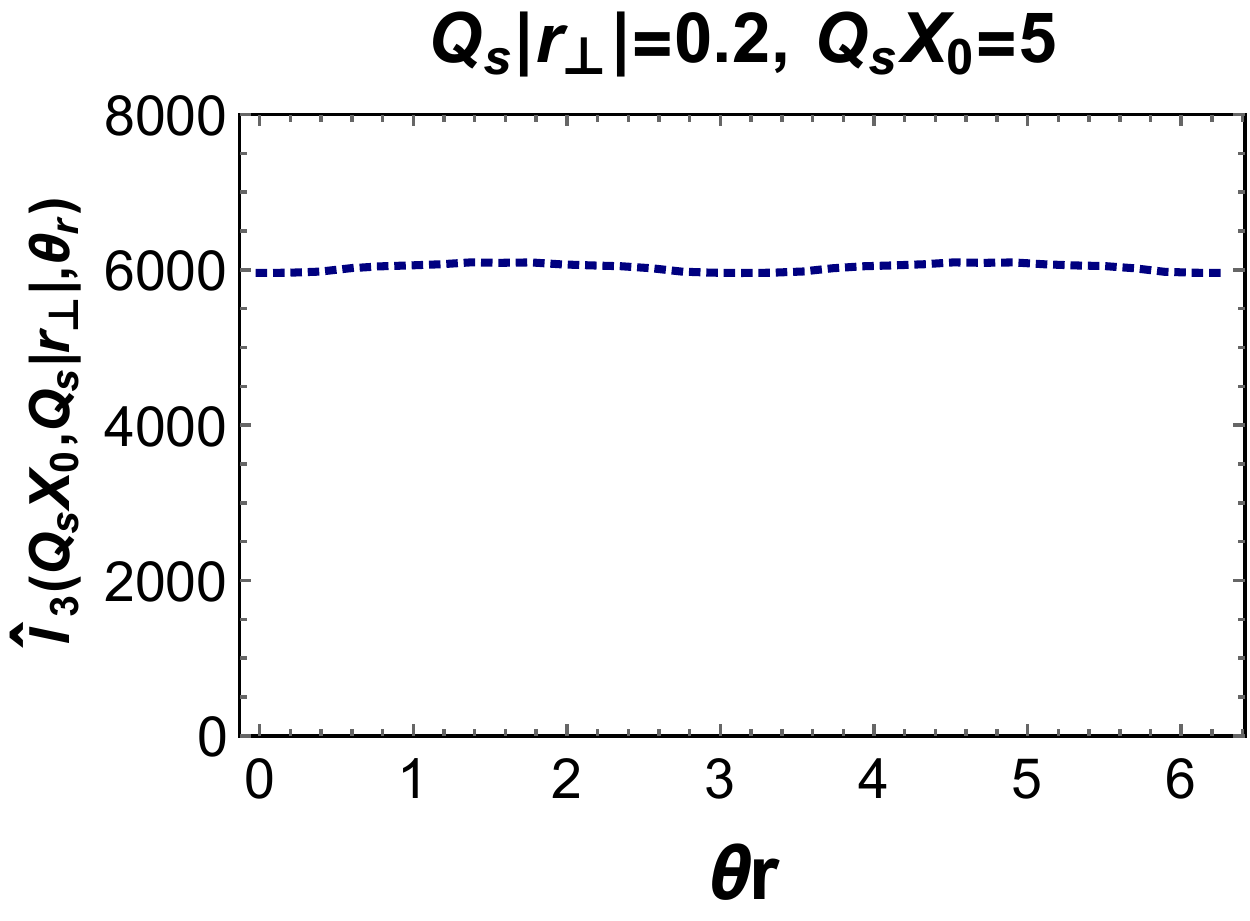}}
			\caption{The $\theta_{r}$ dependence of $\hat{I}_3$ at small $Q_s|r_{\perp}|$.}
			\label{figI3theta}
		\end{center}
	\end{minipage}
\end{figure}

To evaluate $\hat{\mathcal{I}}(Q_sX_0,Q_s|r_{\perp}|,\theta_r)$ numerically\footnote{In principle, we may directly evaluate $\int^{\rm gl} d^2r_{\perp}\mathcal{I}(p_0,Q_sX_0,Q_s|r_{\perp}|)$ numerically. However, to get a rough estimate, we instead adopt the approximation in Eq.~(\ref{eq:Gaussian_approx}) for computational efficiency.}, we utilized the local adaptive method built into $\it{Mathematica}$. As shown in Fig.~\ref{figI123}, it is found that $\hat{\mathcal{I}}_3$ dominates over $\hat{\mathcal{I}}_1$ and $\hat{\mathcal{I}}_2$ particularly at late times $Q_sX_0 >2$ and we may approximate $\hat{\mathcal{I}}\approx \hat{\mathcal{I}}_3$, where $\hat{\mathcal{I}}_{1,2,3}$ correspond to the dimensionless quantities coming from $\mathcal{I}_{1,2,3}$ with the factorization in Eq.~(\ref{eq:I_estimate}). The rapid increase of $\hat{\mathcal{I}}_3(Q_sX_0)$ is predominantly led by the factor $\big(|s'_{\perp}|-|s_{\perp}|\big)\big(|t'_{\perp}|-|t_{\perp}|\big)$ originating from $(X_0''-X_0')(Y_0''-Y_0')$ in the integrand, which can be observed in Fig.~\ref{figI3growth} when comparing with the result by removing this factor. In addition, we have confirmed $\hat{\mathcal{I}}_3(Q_sX_0,Q_s|r_{\perp}|,\pi/2)$ is maximized at $Q_s|r_{\perp}|=0$ as shown in Fig.~\ref{figI3r} with also very mild $\theta_r$ dependence illustrated in Fig.~\ref{figI3theta} at small $Q_s|r_{\perp}|$. One could further verify the same scenario for $\hat{\mathcal{I}}_1(Q_sX_0,Q_s|r_{\perp}|,\theta_r)$ and $\hat{\mathcal{I}}_2(Q_sX_0,Q_s|r_{\perp}|,\theta_r)$. Some qualitative features of $\hat{\mathcal{I}}_{1,2,3}$ are further analyzed in appendix~\ref{app:alternative_I}. 

For convenience, we adopt the approximation in Eq.~(\ref{eq:spin_corr_final}) for conducting numerical estimation. We take the values $Q_s=2$ GeV, $N_c=3$, $A_{\rm N}\sim A_{\rm T}\approx 100$ fm$^{2}$, $m\approx 500$ MeV as the constituent quark mass for strange quarks, and $T\approx 150$ MeV as the freeze-out temperature. By setting $X^{\rm th}_0 \approx 0.5$ fm and correspondingly $Q_sX^{\rm th}_0\approx 5$, which yields $\hat{\mathcal{I}} \approx 6100$ (see Fig.~\ref{figI3growth}), we obtain the result $\langle\mathcal{P}_q^{y}({\bf p})\mathcal{P}_{\bar{q}}^{y}({\bf p})\rangle\approx 0.006$. When choosing a larger saturation scale such as $Q_s=3$ GeV without changing the other parameters, for which now $Q_sX^{\rm th}_0\approx 7.5$ at $X^{\rm th}_0 \approx 0.5$ fm, we obtain $\hat{\mathcal{I}} \approx 20000$ and $\langle\mathcal{P}_q^{y}({\bf p})\mathcal{P}_{\bar{q}}^{y}({\bf p})\rangle\approx 0.05$. Notably, except for the magnification from the prefactor $Q_s^2$, the numerical value of $\hat{\mathcal{I}}(Q_sX^{\rm th}_0)$ also increases with larger $Q_s$ at a fixed thermalization time.

We stress that this estimate is based on rather crude approximations and thus has large uncertainties coming from multiple corrections. Most importantly, we have neglected the spin relaxation in QGP after thermalization and also the distinction between the dressed quark mass in glasma, which should be encoded in $\langle \tilde{a}^{sy}(p,X)\tilde{a}^{sy}(p,Y)\rangle$, and the constituent quark mass for on-shell $f_{\rm eq}(p_0)$. In this scenario, it might be also debatable whether the explicit $1/m^2$ factor in Eq.~(\ref{eq:spin_corr}) for normalization should be set as the constituent quark mass. Both, the spin relaxation time in the strongly coupled QGP and the dressed quark mass in the glasma, are unknown. Fortuitously, the mass-dependent factor $(e^{m/T}+1)^2/m^4$ changes by less than a factor 3 over the range $1 < m/T <4$. 
Despite these uncertainties, we emphasize that these estimated values for spin correlation are substantially larger than the expected vorticity contribution\footnote{From the naive extraction of vorticity contribution to global spin polarization of $\Lambda$ hyperons such that $\mathcal{P}^y_{q}\sim\mathcal{P}^y_{\Lambda}\sim \omega/T\lesssim 10^{-3}$ at high-energy nuclear collisions above 200 GeV \cite{STAR:2017ckg,Becattini:2016gvu} ($\mathcal{P}^y_{\Lambda}\sim 10^{-4}$ at $2.76$ TeV \cite{ALICE:2019onw}), where $\omega$ is the global vorticity, the spin correlation will be approximately $\lesssim 10^{-6}$.}.

\section{Conclusions and outlook}\label{sec:conclusions}
In this paper, we derived the dynamical spin polarization and out-of-plane correlation of massive quarks at small momentum and central rapidity from color fields in the glasma phase in terms of the gluon density. Our formalism is based on quantum kinetic theory (QKT) in the Wigner function formalism. For the GBW distribution, the spin polarization is found to vanish, whereas the out-of-plane spin correlation is non-vanishing and enhanced by $Q_s^2/m^2$ at weak coupling. We have also numerically estimated the magnitude of out-of-plane spin correlation, which is found to be around $0.006\sim 0.05$ at $Q_s=2\sim 3$ GeV and by orders of magnitude larger than the correlation from vorticity for high-energy nuclear collisions.     

As already mentioned in Sec.~\ref{sec:general}, the updated spin coalescence model incorporates both the in-plane (parallel to the reaction plane) and out-of-plane correlations. By symmetry of the color fields from glasma, we may expect $\langle\mathcal{P}_q^{x}({\bf p})\mathcal{P}_{\bar{q}}^{x}({\bf p})\rangle\sim \langle\mathcal{P}_q^{y}({\bf p})\mathcal{P}_{\bar{q}}^{y}({\bf p})\rangle$ at small momenta, while $\langle\mathcal{P}_q^{z}({\bf p})\mathcal{P}_{\bar{q}}^{z}({\bf p})\rangle$ has to be evaluated separately, which remains to be explored for making an estimation of the impact on spin alignment of vector mesons. In addition, it is also important to obtain the momentum dependence for spin correlation and the corresponding spin alignment. Furthermore, the present study only considers the spin correlation between a quark and an antiquark with the same flavor, which is applicable to the case of $\phi$ mesons. For the $K^{*0}$ vector meson, comprising a strange quark and an anti-down quark, further generalization of our formalism for spin correlation is needed. 

Here we used the GBW distribution to evaluate the spin polarization and correlation. Alternatively, we may also apply the gluon distribution \cite{Guerrero-Rodriguez:2021ask} based on the dipole amplitude from the McLerran-Venugopalan model \cite{McLerran:1993ni,McLerran:1993ka,McLerran:1994vd}, which contains a non-vanishing linearly polarized gluon distribution function. It is intriguing to calculate the dynamical spin polarization and correlation with both the unpolarized and linearly polarized gluon distribution functions in such a model using our framework. It is also worth noting that the QKT derived from the $\hbar$ expansion applied to the weak-field limit due to the gradient expansion in phase space. This approximation is similar to the linearization of the Yang-Mills equations by neglecting higher-order effects.  Future studies will be required to construct a self-consistent theory incorporating non-perturbative effects of strong color fields on spin transport. Fortuitously, the nonlinear corrections are more prominent at early times of order $X_0\leq Q_{s}^{-1}$, while the spin correlation is predominantly sensitive to late times. Thus, these higher-order effects may be suppressed.   

\acknowledgments
This work was supported by the National Science and Technology Council (Taiwan) under Grant No. MOST 110-2112-M-001-070-MY3 and by the U.S. Department of Energy under Grant DE-FG02-05ER41367.

\appendix
\section{Spin density matrix}\label{app:density_matrix}
We generalize the derivation in Ref.~\cite{Yang:2017sdk} (as the generalization to incorporate spin degrees of freedom for the coalescence model \cite{Fries:2003vb,Greco:2003xt}) to obtain the spin density matrix when the spin quantization axis does not match the direction for spin polarization of the comprised quark and antiquark in spin-$1$ vector mesons. To construct the spin density matrix of a vector meson from the pair of a quark and an anti-quark through the coalescence model, we may first introduce the density operator of a quark $\rm q$, 
\begin{eqnarray}
	\rho_{\rm q}=V\int \frac{d^3\bm p}{(2\pi)^3}\sum_i\sum_{s_i}w_{\rm q,s_i}|{\rm q};s_{i};\bm p\rangle \langle {\rm q};s_{i};\bm p|,
\end{eqnarray}
where $V$ is the spatial volume and $\bm p$ and $s_i$ denote the spatial momentum and spin, respectively. Here we have to sum over $i=x,y,z$ and $s_i=\pm 1/2$. The spin-dependent weight functions $w_{\rm q,s_i}$ explicitly read
\begin{eqnarray}
	w_{{\rm q},s_i}(\bm p)=\frac{1}{6}+s_i\mathcal{P}_{{\rm q},i}(\bm p),
\end{eqnarray}
where $\mathcal{P}_{{\rm q},i}(\bm p)$ corresponds to the spin polarization of the quark along $i$ direction. Note that the normalization condition 
\begin{eqnarray}
	\sum_i\sum_{s_i}w_{\rm q,s_i}=1
\end{eqnarray}
is satisfied.
Accordingly, 
we can write down the density operator for the pair of a quark and an antiquark via $\rho=\rho_{\rm q}\otimes\rho_{\bar{\rm q}}$, which takes the form
\begin{eqnarray}
	\rho=V^2\int \frac{d^3\bm p}{(2\pi)^3}\int \frac{d^3\bar{\bm p}}{(2\pi)^3}\sum_{i,j}\sum_{s_i,\bar{s}_j}w_{{\rm q},s_i}w_{{\rm q},\bar{s}_j}|{\rm q},\bar{{\rm q}};s_{i},\bar{s}_j;\bm p, \bar{\bm p}\rangle \langle {\rm q},\bar{{\rm q}};s_{i},\bar{s}_j;\bm p, \bar{\bm p}|,
\end{eqnarray}
where 
\begin{eqnarray}
	|{\rm q},\bar{{\rm q}};s_{i},\bar{s}_j;\bm p, \bar{\bm p}\rangle=|{\rm q};s_{i};\bm p\rangle\otimes |\bar{{\rm q}};\bar{s}_j;\bar{\bm p}\rangle.
\end{eqnarray}
For convenience, one could decompose the state of the quark-antiquark pair into the momentum- and spin-dependent parts,
\begin{eqnarray}
	|{\rm q},\bar{{\rm q}};s_{i},\bar{s}_j;\bm p, \bar{\bm p}\rangle=|\bm p, \bar{\bm p}\rangle 	|s_{i},\bar{s}_j\rangle,
\end{eqnarray}
where we omit the quark and antiquark notations for brevity. Also, we take
\begin{eqnarray}
	\langle \bm x,\bar{\bm x}|\bm p, \bar{\bm p}\rangle=V^{-1}e^{i\bm p\cdot \bm x+i\bar{\bm p}\cdot\bar{\bm x}}.
\end{eqnarray} 

Next, we introduce the meson state,
\begin{eqnarray}
	|{\rm M};S,S_z;{\bm P}\rangle=|{\bm P}\rangle|S,S_z\rangle,
\end{eqnarray}
where we have fixed the spin quantization axis along the $z$ direction, which is determined by the experimental setup. Now, $S_z=0$ for $S=0$ and $S_z=\pm 1, 0$ for $S=1$. We also take 
\begin{eqnarray}
	\langle \bm x,\bar{\bm x}|{\bm P}\rangle=V^{-1/2}e^{i\bm P\cdot \bm R}\phi_{M}(\bm y),
\end{eqnarray}
where $\bm R=(\bm x+\bar{\bm x})/2$, $\bm y=\bm x-\bar{\bm x}$, and $\phi_M(\bm y)$ is the normalized meson wave function satisfying $\int d^3\bm y |\phi_{M}(\bm y)|^2=1$.
After choosing the spin quantization axis, one could write down the quark/antiquark spin states explicitly; e.g., 
\begin{eqnarray}
	|s_{x}=\pm 1/2\rangle =\frac{1}{\sqrt{2}}(|+\rangle\pm |-\rangle),\quad |s_{y}=\pm 1/2\rangle =\frac{1}{\sqrt{2}}(|+\rangle \pm i|-\rangle),
\end{eqnarray}
where $|+\rangle=|s_z=1/2\rangle$ and $|-\rangle=|s_z=-1/2\rangle$ and similarly for $|\bar{s}_i\rangle$. The density operator of a quark is accordingly given by  
\begin{eqnarray}
	\rho_{\rm q}=\frac{V}{2}\int \frac{d^3\bm p}{(2\pi)^3}|\bm p\rangle\langle \bm p|
	\begin{pmatrix}\label{eq:dsop_quark}
		1+\mathcal{P}_{{\rm q},z} & \mathcal{P}_{{\rm q},x}-i\mathcal{P}_{{\rm q},y} \\
		\mathcal{P}_{{\rm q},x}+i\mathcal{P}_{{\rm q},y} & 1-\mathcal{P}_{{\rm q},z}
	\end{pmatrix}.
\end{eqnarray}
For $\rho_{\bar{\rm q}}$, one simply has to replace $\bm p$ by $\bar{\bm p}$ and $\mathcal{P}_{{\rm q},i}$ by $\mathcal{P}_{{\bar{\rm q}},i}$ in $\rho_{\rm q}$.
We shall now express the spin states of vector mesons in terms of the bases of the quark-antiquark pair,   
\begin{eqnarray}\nonumber\label{eq:meson_states}
	&&|S=0,S_z=0\rangle=\frac{1}{\sqrt{2}}\big(|+-\rangle -|-+\rangle\big),\quad |S=1,S_z=0\rangle=\frac{1}{\sqrt{2}}\big(|+-\rangle +|-+\rangle\big),
	\\
	&&|S=1,S_z=1\rangle=|++\rangle,\quad |S=1,S_z=-1\rangle=|--\rangle.
\end{eqnarray}
Given Eqs.~(\ref{eq:dsop_quark}) and (\ref{eq:meson_states}), we can now directly evaluate 
\begin{eqnarray}
	\rho^{S}_{S_zS_z}\equiv \langle{\rm M};S,S_z;{\bf P}|\rho|{\rm M};S,S_z;{\bf P}\rangle
\end{eqnarray}
and the normalized spin density matrix for $S=1$,
\begin{eqnarray}
	\rho_{S_zS_z}\equiv \frac{\rho^{S=1}_{S_zS_z}}{\sum_{S_z=\pm 1,0}\rho^{S=1}_{S_zS_z}}.
\end{eqnarray}
Following the calculations in Ref.~\cite{Yang:2017sdk} for handling the states in momentum space, we arrive at
\begin{align}
	\rho_{00}(\bm P)=\frac{\int d^3\bm q \big[1+\sum_{j}\mathcal{P}^{j}_{\rm q}(\bm P/2+\bm q)\mathcal{P}^{j}_{\bar{\rm q}}(\bm P/2-\bm q)-2n\cdot\mathcal{P}_{\rm q}(\bm P/2+\bm q)n\cdot\mathcal{P}_{\bar{\rm q}}(\bm P/2-\bm q)\big]|\tilde{\phi}_{\rm M}(\bm q)|^2}{\int d^3\bm q \big[3+\sum_{j}\mathcal{P}^{j}_{\rm q}(\bm P/2+\bm q)\mathcal{P}^{j}_{\bar{\rm q}}(\bm P/2-\bm q)\big]|\tilde{\phi}_{\rm M}(\bm q)|^2},
\end{align}
where the unit vector $n^{\mu}$ represents the spin quantization axis and 
\begin{eqnarray}
	\tilde{\phi}_{\rm M}(\bm q)\equiv\int d^3\bm y e^{-i\bm q\cdot \bm y}\phi_{\rm M}(\bm y).
\end{eqnarray}
When assuming $|\bm q|\ll |\bm P|$, $\rho_{00}(\bm P)$ reduces to 
\begin{eqnarray}
	\rho_{00}=\frac{1+\sum_{j}\mathcal{P}^{j}_{\rm q}(\bm P/2)\mathcal{P}^{j}_{\bar{\rm q}}(\bm P/2)-2n\cdot\mathcal{P}_{\rm q}(\bm P/2)n\cdot\mathcal{P}_{\bar{\rm q}}(\bm P/2)}{3+\sum_{j}\mathcal{P}^{j}_{\rm q}(\bm P/2)\mathcal{P}^{j}_{\bar{\rm q}}(\bm P/2)}.
\end{eqnarray}

\section{Technical details for the computation of spin correlation}\label{app:spin_corr}
Considering the mid-rapidity region and small-momentum limit, we shall find
\begin{eqnarray}
	\langle{P}^{y}_{q}(p,X){P}^{y}_{q}(p,Y)\rangle
	&\approx&\Bigg(\frac{\tilde{C}_2}{2}
	p_0\big(\partial_{p0}f_V(p_0)\big)\Bigg)^2\int^{p,X}_{k,X'}\int^{p,X'}_{k',X''}\int^{p,Y}_{\bar{k},Y'}\int^{p,Y'}_{\bar{k}',Y''}\nonumber\\&&\Big[\partial_{X''0}\partial_{Y''0}\left\langle E_{[1}^a(X') E_{3]}^a (X'')E_{[1}^b(Y') E_{3]}^b (Y'')\right\rangle +\cdots\Big]
\end{eqnarray}
with the complete expression in Eq.~(\ref{eq:ayay}). Here we simply take one of the terms above as an example.
One may apply the Wick-theorem like decomposition to obtain
\begin{eqnarray}\nonumber
	&&\langle E_{1}^a(X') E_{3}^a (X'')E_{1}^b(Y') E_{3}^b (Y'')\rangle
	\\\nonumber
	&&\approx 
	\langle E_{1}^a(X') E_{3}^a (X'')\rangle\langle E_{1}^b(Y') E_{3}^b (Y'')\rangle
	+\langle E_{1}^a(X')E_{1}^b(Y')\rangle\langle E_{3}^a (X'') E_{3}^b (Y'')\rangle
	\\\nonumber
	&&\quad +\langle E_{1}^a(X') E_{3}^b (Y'')\rangle\langle E_{1}^b(Y') E_{3}^a (X'')\rangle
	\\
	&&=\langle E_{1}^a(X')E_{1}^b(Y')\rangle\langle E_{3}^a (X'') E_{3}^b(Y'')\rangle.
\end{eqnarray} 
To evaluate the multiple integral such as
\begin{eqnarray}
	\mathcal{I}_0(p,X,Y)\equiv \int^{p,X}_{k,X'}\int^{p,X'}_{k',X''}\int^{p,Y}_{\bar{k},Y'}\int^{p,Y'}_{\bar{k}',Y''}
	\partial_{X''0}\partial_{Y''0}\langle E_{1}^a(X')E_{1}^b(Y')\rangle\langle E_{3}^a (X'') E_{3}^b (Y'')\rangle,
\end{eqnarray}
we will use the relation
\begin{eqnarray}\label{eq:useful_int_2}
	\int^{p,X}_{k,X'}G(X,X')\approx\frac{1}{2p_0}\int^{\infty}_{-\infty} d X'_0\Big(1+{\rm sgn}(X_0-X_0')\Big)G(X,X')|_{X'_{1,2,3}=X_{1,2,3}}
\end{eqnarray}
in the small-momentum limit according to Eq.~(\ref{eq:useful_int}). We now have
\begin{eqnarray}\nonumber
	\langle E^{ai}(X')E^{bj}(Y')\rangle&\approx&\frac{1}{2}g^2N_c\delta^{ab}\int \!\frac{d^2q_{\perp}}{(2\pi)^2}\!\!\int \!\frac{d^2l_{\perp}}{(2\pi)^2}\!\!\int \!d^2u_{\perp}\!\!\int \!d^2v_{\perp}
	\\&&\times \Pi_T^{ij}(X'_0,Y'_0,u_{\perp},v_{\perp},q,l)e^{iq_{\perp}(X'-u)_{\perp}}e^{il_{\perp}(Y'-v)_{\perp}},    \label{eq:eiejcorr}
\end{eqnarray}
at small rapidity ($t\approx \tau$), where
\begin{eqnarray}\nonumber
	\Pi_T^{ij}(X'_0,Y'_0,u_{\perp},v_{\perp},q,l)&=&-\epsilon^{in}\epsilon^{jm}\left[G_1(u_{\perp},v_{\perp})G_2(u_{\perp},v_{\perp})-h_1(u_{\perp},v_{\perp})h_2(u_{\perp},v_{\perp})\right]\frac{q^n l^m}{ql}
	\\\nonumber
	&&\times J_1(qX'_0)J_1(lY'_0)\Theta(X'_0)\Theta(Y'_0)
	\\\nonumber
	&=&\frac{q^jl^i-\delta^{ij}{\bm q\cdot\bm l} }{ql}\left[G_1(u_{\perp},v_{\perp})G_2(u_{\perp},v_{\perp})-h_1(u_{\perp},v_{\perp})h_2(u_{\perp},v_{\perp})\right]
	\\
	&&\times J_1(qX'_0)J_1(lY'_0)\Theta(X'_0)\Theta(Y'_0)
	.
\end{eqnarray}
Similarly, we have 
\begin{eqnarray}\nonumber
	\langle E^{a}_3(X'')E^{b}_3(Y'')\rangle&\approx&\frac{1}{2}g^2N_c\delta^{ab}\!\!\int \!\frac{d^2q_{\perp}}{(2\pi)^2}\!\!\int \!\frac{d^2l_{\perp}}{(2\pi)^2}\!\!\int \!d^2u_{\perp}\!\!\int \!d^2v_{\perp}
	\\
	&&	\times \Pi_L(X''_0,Y''_0,u_{\perp},v_{\perp},q,l)e^{iq_{\perp}(X'-u)_{\perp}}e^{il_{\perp}(Y'-v)_{\perp}},
\end{eqnarray}
where
\begin{eqnarray}\nonumber
	\Pi_L(X'_0,Y'_0,u_{\perp},v_{\perp},q,l)&=&\left[G_1(u_{\perp},v_{\perp})G_2(u_{\perp},v_{\perp})+h_1(u_{\perp},v_{\perp})h_2(u_{\perp},v_{\perp})\right]
	\\
	&&\times J_0(qX''_0)J_0(lY''_0)\Theta(X''_0)\Theta(Y''_0).
\end{eqnarray}

It turns out that
\begin{eqnarray}\nonumber\label{eq:IpXY}
	\mathcal{I}_0(p,X,Y)&\approx& \frac{g^4N_c^2(N_c^2-1)}{4}\left(\frac{1}{2p_0}\right)^4
	\int^{X_0}_{X'_0}\int^{Y_0}_{Y'_0}\int^{X'_0}_{X''_0}\int^{Y'_0}_{Y''_0}
	\\\nonumber
	&&\int \!\frac{d^2q_{\perp}}{(2\pi)^2}\!\!\int \!\frac{d^2l_{\perp}}{(2\pi)^2}\!\!\int \!d^2u_{\perp}\!\!\int \!d^2v_{\perp}
	\int \!\frac{d^2q'_{\perp}}{(2\pi)^2}\!\!\int \!\frac{d^2l'_{\perp}}{(2\pi)^2}\!\!\int \!d^2u'_{\perp}\!\!\int \!d^2v'_{\perp}
	\\\nonumber
	&&\times e^{iq_{\perp}(X-u)_{\perp}}e^{il_{\perp}(Y-v)_{\perp}}
	e^{iq'_{\perp}(X-u')_{\perp}}e^{il'_{\perp}(Y-v')_{\perp}}
	\\
	&&\times\Pi_T^{xx}(X'_0,Y'_0,u_{\perp},v_{\perp},q,l)\partial_{X''0}\partial_{Y''0}\Pi_L(X''_0,Y''_0,u'_{\perp},v'_{\perp},q',l')
\end{eqnarray}
by using Eq.~(\ref{eq:useful_int_2}),
where
\begin{eqnarray}
	\int^{X_0}_{X'_0}\equiv \int^{\infty}_{-\infty} d X'_0\Big(1+{\rm sgn}(X_0-X_0')\Big).
\end{eqnarray}
On can rewrite Eq.~(\ref{eq:IpXY}) into a more compact form as
\begin{eqnarray}\nonumber\label{eq:IpXY1}
	\mathcal{I}_0(p,X,Y)&\approx& \frac{g^4N_c^2(N_c^2-1)}{64p_0^4}
	\int \!\frac{d^2q_{\perp}}{(2\pi)^2}\!\!\int \!\frac{d^2l_{\perp}}{(2\pi)^2}\!\!\int \!d^2u_{\perp}\!\!\int \!d^2v_{\perp}
	\int \!\frac{d^2q'_{\perp}}{(2\pi)^2}\!\!\int \!\frac{d^2l'_{\perp}}{(2\pi)^2}\!\!\int \!d^2u'_{\perp}\!\!\int \!d^2v'_{\perp}
	\\\nonumber
	&&\times e^{iq_{\perp}(X-u)_{\perp}}e^{il_{\perp}(Y-v)_{\perp}}
	e^{iq'_{\perp}(X-u')_{\perp}}e^{il'_{\perp}(Y-v')_{\perp}}
	\\
	&&\times \left(\frac{-q^yl^y}{q l}\right)\Omega_{-}(u_{\perp},v_{\perp})\Omega_{+}(u'_{\perp},v'_{\perp})\mathcal{Y}(X_0,Y_0,q,l,q',l'),
\end{eqnarray}
where
\begin{eqnarray}
	&&\mathcal{Y}(X_0,Y_0,q,l,q',l')
	\\\nonumber
	&&\equiv\int^{X_0}_{X'_0}\int^{Y_0}_{Y'_0}\int^{X'_0}_{X''_0}\int^{Y'_0}_{Y''_0}
	J_1(qX'_0)J_1(lY'_0)\Theta(X'_0)\Theta(Y'_0)\partial_{X''0}\partial_{Y''0}
	J_0(q'X''_0)J_0(l'Y''_0)\Theta(X''_0)\Theta(Y''_0).
\end{eqnarray}
We may first evaluate $\mathcal{Y}(X_0,Y_0,q,l,q',l')$. Using the integration by part and $\partial_x{\rm sgn}(x)=2\delta(x)$, we obtain
\begin{eqnarray}\nonumber
	\int^{Y'_0}_{Y''_0}\partial_{Y''0}
	J_0(l'Y''_0)\Theta(Y''_0)&=&(1+{\rm sgn}(Y_0'-Y_0''))J_0(l'Y''_0)\Theta(Y''_0)\big|^{Y'_0=\infty}_{{Y'_0=-\infty}}
	\\\nonumber
	&&+2\int^{\infty}_{\infty}\delta(Y_0'-Y_0'')J_0(l'Y''_0)\Theta(Y''_0)
	\\
	&=&2J_0(l'Y'_0)\Theta(Y'_0).
\end{eqnarray}
It is hence found
\begin{eqnarray}
	\int^{X'_0}_{X''_0}\int^{Y'_0}_{Y''_0}\partial_{X''0}\partial_{Y''0}
	J_0(q'X''_0)J_0(l'Y''_0)\Theta(X''_0)\Theta(Y''_0)=4J_0(q'X'_0)\Theta(X'_0)J_0(l'Y'_0)\Theta(Y'_0)
\end{eqnarray}
and accordingly
\begin{eqnarray}\nonumber
	\mathcal{Y}(X_0,Y_0,q,l,q',l')
	&=&16\int^{\infty}_{-\infty} dX_0'\int^{\infty}_{-\infty} dY_0'
	J_1(qX'_0)J_1(lY'_0)J_0(q'X'_0)J_0(l'Y'_0)
	\\
	&&\times \Theta(X_0-X'_0)\Theta(Y_0-Y'_0)\Theta(X'_0)\Theta(Y'_0),
\end{eqnarray}
where we have utilized $\Theta(x){\rm sgn}(x)=\Theta(x)$ and $(1+{\rm sgn}(x_0-x))\Theta(x)=2\Theta(x_0-x)\Theta(x)$.

Next, we will consider the integration over $q'$ and $l'$. It will be more convenient to make the decomposition,
\begin{eqnarray}\label{eq:decomp_qi}
	q^i=\frac{(X-u)_{\perp}^i}{|X_{\perp}-u_{\perp}|}q\cos\theta_q+\Theta^{ij}_{X-u}q_j\sin\theta_q,\quad
	l^i=\frac{(Y-v)_{\perp}^i}{|Y_{\perp}-v_{\perp}|}l\cos\theta_l+\Theta^{ij}_{Y-v}l_j\sin\theta_l,
\end{eqnarray} 
where $\Theta^{ij}_{V}\equiv \eta^{ij}_{\perp}+V^{i}_{\perp}V^{j}_{\perp}/|V_{\perp}|^2$. Here $\theta_q$ and $\theta_l$ should appear in $\int d^2q_{\perp}=\int dqqd\theta_{q}$ and $\int d^2l_{\perp}=\int dlld\theta_{l}$. We can accordingly evaluate the related integral
\begin{eqnarray}
	&&\int d\theta_q\int d\theta_{l}e^{iq_{\perp}(X'-u)_{\perp}}e^{il_{\perp}(Y'-v)_{\perp}}=(2\pi)^2J_0(q|X_{\perp}-u_{\perp}|)J_0(l|Y_{\perp}-v_{\perp}|),
	\\
	\nonumber
	&&\int d\theta_q\int d\theta_{l}\frac{q^jl^i-\delta^{ij}{\bm q\cdot\bm l} }{ql}e^{iq_{\perp}(X'-u)_{\perp}}e^{il_{\perp}(Y'-v)_{\perp}}
	\\
	&&=(2\pi)^2\frac{\delta^{ij}(\bm X-\bm u)_{\perp}\cdot (\bm Y-\bm v)_{\perp}-(X-u)_{\perp}^i(Y-v)_{\perp}^j}{|X_{\perp}-u_{\perp}||Y_{\perp}-v_{\perp}|}J_1(q|X_{\perp}-u_{\perp}|)J_1(l|Y_{\perp}-v_{\perp}|),
\end{eqnarray}
by using
\begin{eqnarray}\nonumber\label{eq:thetaint_1}
	&&\int^{2\pi}_0 d\theta e^{ia\cos \theta}=2\pi J_0(|a|),\quad \int^{2\pi}_0 d\theta e^{ia\cos \theta}\cos(\theta+b)=2i\pi J_1(a)\cos b,
	\\
	&&\int^{2\pi}_0 d\theta e^{ia\cos \theta}\sin(\theta+b)=2i\pi J_1(a)\sin b.
\end{eqnarray}
Note that $(\bm X-\bm u)_{\perp}\cdot (\bm X-\bm v)_{\perp}$ can be written as $(\bm X-\bm u)_{\perp}\cdot (\bm X-\bm v)_{\perp}=|X_{\perp}-u_{\perp}||X_{\perp}-v_{\perp}|\cos(\theta_{X-u}-\theta_{X-v})$. 

One then obtains
\begin{eqnarray}\nonumber\label{eq:IpXY2}
	\mathcal{I}_0(p,X,Y)&\approx& \frac{g^4N_c^2(N_c^2-1)}{64p_0^4}
	\int \!\frac{dqq}{(2\pi)^2}\!\!\int \!dll\!\!\int \!d^2u_{\perp}\!\!\int \!d^2v_{\perp}
	\int \!\frac{dq'q'}{(2\pi)^2}\!\!\int \!dl' l'\!\!\int \!d^2u'_{\perp}\!\!\int \!d^2v'_{\perp}
	\\\nonumber
	&&\times J_1(q|X_{\perp}-u_{\perp}|)J_1(l|Y_{\perp}-v_{\perp}|) J_0(q'|X_{\perp}-u'_{\perp}|)J_0(l'|Y_{\perp}-v'_{\perp}|)
	\\
	&&\times \left(\frac{(X-u)_{\perp}^x(Y-v)_{\perp}^x}{|X_{\perp}-u_{\perp}||Y_{\perp}-v_{\perp}|}\right)\Omega_{-}(u_{\perp},v_{\perp})\Omega_{+}(u'_{\perp},v'_{\perp})\mathcal{Y}(X_0,Y_0,q,l,q',l').
\end{eqnarray}
Using Eq.~(\ref{eq:Bessel_rel}), we acquire
\begin{eqnarray}\nonumber
	&&\int \!\frac{dq'q'}{(2\pi)^2}\!\!\int \!dl' l'\mathcal{Y}(X_0,Y_0,q,l,q',l')J_0(q'|X_{\perp}-u'_{\perp}|)J_0(l'|Y_{\perp}-v'_{\perp}|)
	\\\nonumber
	&&=\frac{16}{(2\pi)^2}\int^{\infty}_{-\infty} dX_0'\int^{\infty}_{-\infty} dY_0'
	J_1(qX'_0)J_1(lY'_0)
	\Theta(X_0-X'_0)\Theta(Y_0-Y'_0)\Theta(X'_0)\Theta(Y'_0)
	\\\nonumber
	&&\quad\times  \frac{\delta(X_0'-|X_{\perp}-u'_{\perp}|)}{|X_{\perp}-u'_{\perp}|}\frac{\delta(Y_0'-|Y_{\perp}-v'_{\perp}|)}{|Y_{\perp}-v'_{\perp}|}
	\\
	&&=\frac{16J_1(q|X_{\perp}-u'_{\perp}|)J_1(l|Y_{\perp}-v'_{\perp}|)}{(2\pi)^2|X_{\perp}-u'_{\perp}||Y_{\perp}-v'_{\perp}|}\Theta(X_0-|X_{\perp}-u'_{\perp}|)\Theta(Y_0-|Y_{\perp}-v'_{\perp}|)
	.
\end{eqnarray}
Subsequently, taking for example
\begin{eqnarray}
	\int dq qJ_1(q|X_{\perp}-u_{\perp}|)J_1(q|X_{\perp}-u'_{\perp}|)=\frac{\delta(|X_{\perp}-u_{\perp}|-|X_{\perp}-u'_{\perp}|)}{|X_{\perp}-u_{\perp}|},
\end{eqnarray}
one finds
\begin{eqnarray}\nonumber
	\mathcal{I}_0(p,X,Y)&\approx& \frac{g^4N_c^2(N_c^2-1)}{4(2\pi)^4p_0^4}\int^{\perp}_{u,v,u'v'}\left(\frac{(X-u)_{\perp}^x(Y-v)_{\perp}^x}{|X_{\perp}-u_{\perp}||Y_{\perp}-v_{\perp}|}\right)\Omega_{-}(u_{\perp},v_{\perp})\Omega_{+}(u'_{\perp},v'_{\perp})
	\\\nonumber
	&&\times \frac{\delta(|X_{\perp}-u_{\perp}|-|X_{\perp}-u'_{\perp}|)}{|X_{\perp}-u_{\perp}|^2}
	\frac{\delta(|Y_{\perp}-v_{\perp}|-|Y_{\perp}-v'_{\perp}|)}{|Y_{\perp}-v_{\perp}|^2}
	\\
	&&\times \Theta(X_0-|X_{\perp}-u_{\perp}|)\Theta(Y_0-|Y_{\perp}-v_{\perp}|).
\end{eqnarray}
Given explicit expressions of $\Omega_{\pm}$ from Ref.~\cite{Guerrero-Rodriguez:2021ask}, one can evaluate $\mathcal{I}_0(p,X,Y)$ numerically.

\section{Detailed proof for the vanishing term}\label{app:vanishing_terms}
Considering the first term in (\ref{eq:ayay}), by symmetry, on finds
\begin{eqnarray}\nonumber
	&&\int^{p,X}_{k,X'}\int^{p,X'}_{k',X''}\int^{p,Y}_{\bar{k},Y'}\int^{p,Y'}_{\bar{k}',Y''}\partial_{X''0}\partial_{Y''0}\left\langle E_{[1}^a(X') E_{3]}^a (X'')E_{[1}^b(Y') E_{3]}^b (Y'')\right\rangle
	\\\nonumber
	&&=\int^{p,X}_{k,X'}\int^{p,X'}_{k',X''}\int^{p,Y}_{\bar{k},Y'}\int^{p,Y'}_{\bar{k}',Y''}\partial_{X''0}\partial_{Y''0}\Big(\left\langle E_{1}^a(X') E_{3}^a (X'')E_{1}^b(Y') E_{3}^b (Y'')\right\rangle
	\\
	&&\quad +\left\langle E_{3}^a(X') E_{1}^a (X'')E_{3}^b(Y') E_{1}^b (Y'')\right\rangle
	-2\left\langle E_{1}^a(X') E_{3}^a (X'')E_{3}^b(Y') E_{1}^b (Y'')\right\rangle\Big).
\end{eqnarray}
Since 
\begin{eqnarray}
	\left\langle E_{3}^a(X') E_{1}^a (X'')E_{3}^b(Y') E_{1}^b (Y'')\right\rangle&=&\left\langle E_{1}^a(X') E_{3}^a (X'')E_{1}^b(Y') E_{3}^b (Y'')\right\rangle|_{X'\leftrightarrow X'',Y'\leftrightarrow Y''},
	\\
	\left\langle E_{1}^a(X') E_{3}^a (X'')E_{3}^b(Y') E_{1}^b (Y'')\right\rangle&=&\left\langle E_{1}^a(X') E_{3}^a (X'')E_{1}^b(Y') E_{3}^b (Y'')\right\rangle|_{Y'\leftrightarrow Y''},
\end{eqnarray}
when $X'_{1,2,3}=X''_{1,2,3}=X_{1,2,3}$ and $Y'_{1,2,3}=Y''_{1,2,3}=Y_{1,2,3}$, based on Eq.~(\ref{eq:IpXY2}), one immediately finds
\begin{eqnarray}\nonumber
	&&\int^{p,X}_{k,X'}\int^{p,X'}_{k',X''}\int^{p,Y}_{\bar{k},Y'}\int^{p,Y'}_{\bar{k}',Y''}\partial_{X''0}\left\langle E_{3}^a(X') E_{1}^a (X'')E_{3}^b(Y') E_{1}^b (Y'')\right\rangle
	\\\nonumber
	&&\approx\frac{g^4N_c^2(N_c^2-1)}{64p_0^4}
	\int \!\frac{dqq}{(2\pi)^2}\!\!\int \!dll\!\!\int \!d^2u_{\perp}\!\!\int \!d^2v_{\perp}
	\int \!\frac{dq'q'}{(2\pi)^2}\!\!\int \!dl' l'\!\!\int \!d^2u'_{\perp}\!\!\int \!d^2v'_{\perp}
	\\\nonumber
	&&\times J_1(q|X_{\perp}-u_{\perp}|)J_1(l|Y_{\perp}-v_{\perp}|) J_0(q'|X_{\perp}-u'_{\perp}|)J_0(l'|Y_{\perp}-v'_{\perp}|)
	\\
	&&\times \left(\frac{(X-u)_{\perp}^x(Y-v)_{\perp}^x}{|X_{\perp}-u_{\perp}||Y_{\perp}-v_{\perp}|}\right)\Omega_{-}(u_{\perp},v_{\perp})\Omega_{+}(u'_{\perp},v'_{\perp})\mathcal{Y}_1(X_0,Y_0,q,l,q',l'),
\end{eqnarray}
where 
\begin{eqnarray}\nonumber
	&&\mathcal{Y}_1(X_0,Y_0,q,l,q',l')
	\\\nonumber
	&&=\int^{X_0}_{X'_0}\int^{Y_0}_{Y'_0}\int^{X'_0}_{X''_0}\int^{Y'_0}_{Y''_0}\partial_{X''0}\partial_{Y''0}
	J_1(qX''_0)J_1(lY''_0)
	J_0(q'X'_0)J_0(l'Y'_0)\Theta(X'_0)\Theta(Y'_0)\Theta(X''_0)\Theta(Y''_0)
	\\\nonumber
	&&=16\int^{X_0}_0dX'_0\int^{Y_0}_0dY'_0J_1(qX'_0)J_1(lY'_0)
	J_0(q'X'_0)J_0(l'Y'_0)
	\\
	&&=\mathcal{Y}(X_0,Y_0,q,l,q',l').
\end{eqnarray}
Similarly, it is found
\begin{eqnarray}\nonumber
	&&\int^{p,X}_{k,X'}\int^{p,X'}_{k',X''}\int^{p,Y}_{\bar{k},Y'}\int^{p,Y'}_{\bar{k}',Y''}\partial_{X''0}\left\langle E_{1}^a(X') E_{3}^a (X'')E_{3}^b(Y') E_{1}^b(Y'')\right\rangle
	\\\nonumber
	&&\approx\frac{g^4N_c^2(N_c^2-1)}{64p_0^4}
	\int \!\frac{dqq}{(2\pi)^2}\!\!\int \!dll\!\!\int \!d^2u_{\perp}\!\!\int \!d^2v_{\perp}
	\int \!\frac{dq'q'}{(2\pi)^2}\!\!\int \!dl' l'\!\!\int \!d^2u'_{\perp}\!\!\int \!d^2v'_{\perp}
	\\\nonumber
	&&\times J_1(q|X_{\perp}-u_{\perp}|)J_1(l|Y_{\perp}-v_{\perp}|) J_0(q'|X_{\perp}-u'_{\perp}|)J_0(l'|Y_{\perp}-v'_{\perp}|)
	\\
	&&\times \left(\frac{(X-u)_{\perp}^x(Y-v)_{\perp}^x}{|X_{\perp}-u_{\perp}||Y_{\perp}-v_{\perp}|}\right)\Omega_{-}(u_{\perp},v_{\perp})\Omega_{+}(u'_{\perp},v'_{\perp})\mathcal{Y}_2(X_0,Y_0,q,l,q',l'),
\end{eqnarray}
where
\begin{eqnarray}\nonumber
	&&\mathcal{Y}_2(X_0,Y_0,q,l,q',l')
	\\\nonumber
	&&=\int^{X_0}_{X'_0}\int^{Y_0}_{Y'_0}\int^{X'_0}_{X''_0}\int^{Y'_0}_{Y''_0}\partial_{X''0}\partial_{Y''0}
	J_1(qX'_0)J_1(lY''_0)
	J_0(q'X''_0)J_0(l'Y'_0)\Theta(X'_0)\Theta(Y'_0)\Theta(X''_0)\Theta(Y''_0)
	\\\nonumber
	&&=16\int^{X_0}_0dX'_0\int^{Y_0}_0dY'_0J_1(qX'_0)J_1(lY'_0)
	J_0(q'X'_0)J_0(l'Y'_0)
	\\
	&&=\mathcal{Y}(X_0,Y_0,q,l,q',l').
\end{eqnarray}
It turns out that
\begin{eqnarray}
	\int^{p,X}_{k,X'}\int^{p,X'}_{k',X''}\int^{p,Y}_{\bar{k},Y'}\int^{p,Y'}_{\bar{k}',Y''}\partial_{X''0}\partial_{Y''0}\left\langle E_{[1}^a(X') E_{3]}^a (X'')E_{[1}^b(Y') E_{3]}^b (Y'')\right\rangle=0.
\end{eqnarray}
In light of the same approach, all the terms associated with $\partial_{X0''} E_{[1}^a(X') E_{3]}^a (X'')$ in Eq.~(\ref{eq:ayay}) vanish simply due to 
\begin{eqnarray}
	\int^{X'_0}_{X_0''}\partial_{X0''}J_1(qX_0'')J_0(lX'_0)\Theta(X_0')\Theta(X_0'')=
	\int^{X'_0}_{X_0''}\partial_{X0''}J_1(qX_0')J_0(lX''_0)\Theta(X_0')\Theta(X_0'').
\end{eqnarray}
.

\section{Remaining terms}\label{app:remaining_terms}
We may now consider
\begin{eqnarray}\nonumber
	\mathcal{I}_{2}&\equiv& -2\int^{p,X}_{k,X'}\int^{p,X'}_{k',X''}\int^{p,Y}_{\bar{k},Y'}\int^{p,Y'}_{\bar{k}',Y''}(Y_0''-Y_0')\Big[\partial_{X'1}\partial^2_{Y''1}\left\langle B^{a[2}(X')E^{a1]}(X'')E_1^b(Y') E_3^b (Y'')\right\rangle
	\\\nonumber
	&&+\partial_{X'1}\partial_{Y''2}\partial_{Y''1}\left\langle B^{a[2}(X')E^{a1]}(X'')E_2^b(Y') E_3^b (Y'')\right\rangle\Big]
	\\\nonumber
	&\approx& \frac{-1}{8p_0^4}\int^{X_0}_{X'_0}\int^{X'_0}_{X''_0}\int^{Y_0}_{Y'_0}\int^{Y'_0}_{Y''_0}(Y_0''-Y_0')\Big[\partial_{X'1}\partial^2_{Y''1}\left\langle B^{a[2}(X')E^{a1]}(X'')E_1^b(Y') E_3^b (Y'')\right\rangle
	\\
	&&+\partial_{X'1}\partial_{Y''2}\partial_{Y''1}\left\langle B^{a[2}(X')E^{a1]}(X'')E_2^b(Y') E_3^b (Y'')\right\rangle\Big]
\end{eqnarray}
with $X_{\perp}=X'_{\perp}=X''_{\perp}$ and $Y_{\perp}=Y'_{\perp}=Y''_{\perp}$.
By using
\begin{eqnarray}\nonumber
	\left\langle B^{a[2}(X')E^{a1]}(X'')E_1^b(Y') E_3^b (Y'')\right\rangle
	&=&\langle E^{b1}(Y')E^{a[1}(X'')\rangle\langle B^{a2]}(X')E^{b3}(Y'')\rangle
	\\\nonumber
	&=&-i\bar{N}_c^2(N_c^2-1)\int^{X'}_{\perp;q,u}\int^{Y'}_{\perp;l,v}\int^{X''}_{\perp;q',u'}\int^{Y''}_{\perp;l',v'}
	\Omega_{-}(v_{\perp},u'_{\perp})\Omega_{+}(u_{\perp},v'_{\perp})
	\\
	&&
	\times\frac{q^{[x} q'^{y]}l^y}{qq'l}
	\!J_1(lY'_0)J_1(q'X''_0)J_1(qX'_0)J_0(l'Y''_0)
\end{eqnarray}
and
\begin{eqnarray}\nonumber
	\left\langle B^{a[2}(X')E^{a1]}(X'')E_2^b(Y') E_3^b (Y'')\right\rangle&=&\langle E^{b2}(Y')E^{a[1}(X'')\rangle\langle B^{a2]}(X')E^{b3}(Y'')\rangle
	\\\nonumber
	&=&i\bar{N}_c^2(N_c^2-1)\int^{X'}_{\perp;q,u}\int^{Y'}_{\perp;l,v}\int^{X''}_{\perp;q',u'}\int^{Y''}_{\perp;l',v'}
	\Omega_{-}(v_{\perp},u'_{\perp})\Omega_{+}(u_{\perp},v'_{\perp})
	\\
	&&
	\times\frac{q^{[x} q'^{y]}l^x}{qq'l}
	\!J_1(lY'_0)J_1(q'X''_0)J_1(qX'_0)J_0(l'Y''_0),
\end{eqnarray} 
one obtains
\begin{eqnarray}\nonumber
	\mathcal{I}_{2}
	&\approx& \frac{i\bar{N}_c^2(N_c^2-1)}{8p_0^4}\int^{X_0}_{X'_0}\int^{X'_0}_{X''_0}\int^{Y_0}_{Y'_0}\int^{Y'_0}_{Y''_0}\int^{X}_{\perp;q,u}\int^{Y}_{\perp;l,v}\int^{X}_{\perp;q',u'}\int^{Y}_{\perp;l',v'}\!J_1(lY'_0)J_1(q'X''_0)J_1(qX'_0)J_0(l'Y''_0)
	\\
	&&\times(Y_0''-Y_0')\frac{q^{[x} q'^{y]}(l^y\partial_{v'x}-l^x\partial_{v'y})}{qq'l}\partial_{ux}\partial_{v'x}
	\Omega_{-}(v_{\perp},u'_{\perp})\Omega_{+}(u_{\perp},v'_{\perp}).
\end{eqnarray}
Then, implementing Eq.~(\ref{eq:decomp_qi}), Eq.~(\ref{eq:thetaint_1}), and Eq.~(\ref{eq:Bessel_rel}) yields
\begin{eqnarray}\nonumber
\mathcal{I}_{2}
&\approx& \frac{g^4N_c^2(N_c^2-1)}{2(2\pi)^4p_0^4}
\int^{\perp}_{s,t,s',t'}\frac{\Theta(X_0-|s_{\perp}|)\Theta(Y_0-|t_{\perp}|)\Theta(|s_{\perp}|-|s'_{\perp}|)\Theta(|t_{\perp}|-|t'_{\perp}|)}{|s_{\perp}||t_{\perp}||s'_{\perp}||t'_{\perp}|}
\\
&&\times\big(|t'_{\perp}|-|t_{\perp}|\big)\hat{s}_{\perp}^{[x} \hat{s}'^{ y]}_{\perp}\big(\hat{t}^y_{\perp}\partial_{v'x}-\hat{t}^x_{\perp}\partial_{v'y}\big)\partial_{ux}\partial_{v'x}
\Omega_{-}(v_{\perp},u'_{\perp})\Omega_{+}(u_{\perp},v'_{\perp}).
\end{eqnarray}
For convenience, one may adopt the change of variables, $v_{\perp}\leftrightarrow v'_{\perp}$, which yields 
\begin{eqnarray}\nonumber
	\mathcal{I}_{2}
	&\approx& \frac{g^4N_c^2(N_c^2-1)}{2(2\pi)^4p_0^4}
	\int^{\perp}_{s,t,s',t'}\frac{\Theta(X_0-|s_{\perp}|)\Theta(Y_0-|t_{\perp}|)\Theta(|s_{\perp}|-|s'_{\perp}|)\Theta(|t'_{\perp}|-|t_{\perp}|)}{|s_{\perp}||t_{\perp}||s'_{\perp}||t'_{\perp}|}
	\\
	&&\times\big(|t_{\perp}|-|t'_{\perp}|\big)\hat{s}_{\perp}^{[x} \hat{s}'^{ y]}_{\perp}\big(\hat{t}'^y_{\perp}\partial_{vx}-\hat{t}'^x_{\perp}\partial_{vy}\big)\partial_{ux}\partial_{vx}
	\Omega_{-}(v'_{\perp},u'_{\perp})\Omega_{+}(u_{\perp},v_{\perp}).
\end{eqnarray}

Finally, we would like to evaluate
\begin{eqnarray}\nonumber
	\mathcal{I}_{3}&\equiv& \int^{p,X}_{k,X'}\int^{p,X'}_{k',X''}\int^{p,Y}_{\bar{k},Y'}\int^{p,Y'}_{\bar{k}',Y''}(X_0''-X_0')(Y_0''-Y_0')\Big(\partial^2_{X''1}\partial^2_{Y''1}\left\langle E_1^a(X') E_3^a (X'')E_{1}^b(Y') E_{3}^b (Y'')\right\rangle\nonumber\\
	&&+2\partial^2_{X''1}\partial_{Y''2}\partial_{Y''1}\left\langle E_1^a(X') E_3^a (X'')E_2^b(Y') E_3^b (Y'')\right\rangle\nonumber\\
	&&+\partial_{X''2}\partial_{X''1}\partial_{Y''2}\partial_{Y''1}\left\langle E_2^a(X') E_3^a (X'')E_2^b(Y') E_3^b (Y'')\right\rangle\Big)
	\nonumber
	\\\nonumber
	&\approx& \frac{1}{16p_0^4}\int^{X_0}_{X'_0}\int^{X'_0}_{X''_0}\int^{Y_0}_{Y'_0}\int^{Y'_0}_{Y''_0}(X_0''-X_0')(Y_0''-Y_0')\Big(\partial^2_{X''1}\partial^2_{Y''1}\left\langle E_1^a(X') E_3^a (X'')E_{1}^b(Y') E_{3}^b (Y'')\right\rangle\nonumber\\
	&&+2\partial^2_{X''1}\partial_{Y''2}\partial_{Y''1}\left\langle E_1^a(X') E_3^a (X'')E_2^b(Y') E_3^b (Y'')\right\rangle\nonumber\\
	&&+\partial_{X''2}\partial_{X''1}\partial_{Y''2}\partial_{Y''1}\left\langle E_2^a(X') E_3^a (X'')E_2^b(Y') E_3^b (Y'')\right\rangle\Big)
\end{eqnarray}
with $X_{\perp}=X'_{\perp}=X''_{\perp}$ and $Y_{\perp}=Y'_{\perp}=Y''_{\perp}$. Using
\begin{eqnarray}\nonumber
\left\langle E_1^a(X') E_3^a (X'')E_{1}^b(Y') E_{3}^b (Y'')\right\rangle&=&\langle E_1^a(X')E_{1}^b(Y')\rangle \langle E_3^a (X'') E_{3}^b (Y'')\rangle
\\\nonumber
&=&-\bar{N}_c^2(N_c^2-1)\int^{X'}_{\perp;q,u}\int^{Y'}_{\perp;l,v}\int^{X''}_{\perp;q',u'}\int^{Y''}_{\perp;l',v'}
\Omega_{-}(u_{\perp},v_{\perp})\Omega_{+}(u'_{\perp},v'_{\perp})
\\
&&\times\frac{q^y l^y}{ql}
\!J_1(qX'_0)J_1(lY'_0)J_0(q'X''_0)J_0(l'Y''_0),
\end{eqnarray}
\begin{eqnarray}\nonumber
\left\langle E_1^a(X') E_3^a (X'')E_2^b(Y') E_3^b (Y'')\right\rangle&=&\langle E_1^a(X')E_{2}^b(Y')\rangle \langle E_3^a (X'') E_{3}^b (Y'')\rangle
\\\nonumber
&=&\bar{N}_c^2(N_c^2-1)\int^{X'}_{\perp;q,u}\int^{Y'}_{\perp;l,v}\int^{X''}_{\perp;q',u'}\int^{Y''}_{\perp;l',v'}
\Omega_{-}(u_{\perp},v_{\perp})\Omega_{+}(u'_{\perp},v'_{\perp})
\\
&&\times\frac{q^y l^x}{ql}
\!J_1(qX'_0)J_1(lY'_0)J_0(q'X''_0)J_0(l'Y''_0),
\end{eqnarray}
and
\begin{eqnarray}\nonumber
\left\langle E_2^a(X') E_3^a (X'')E_2^b(Y') E_3^b (Y'')\right\rangle&=&\langle E_2^a(X')E_{2}^b(Y')\rangle \langle E_3^a (X'') E_{3}^b (Y'')\rangle
\\\nonumber
&=&-\bar{N}_c^2(N_c^2-1)\int^{X'}_{\perp;q,u}\int^{Y'}_{\perp;l,v}\int^{X''}_{\perp;q',u'}\int^{Y''}_{\perp;l',v'}
\Omega_{-}(u_{\perp},v_{\perp})\Omega_{+}(u'_{\perp},v'_{\perp})
\\
&&\times\frac{q^x l^x}{ql}
\!J_1(qX'_0)J_1(lY'_0)J_0(q'X''_0)J_0(l'Y''_0),
\end{eqnarray}
and following the same procedure, it is found
\begin{eqnarray}\nonumber
	\mathcal{I}_{3}
	&\approx& -\frac{g^4N_c^2(N_c^2-1)}{4(2\pi)^4p_0^4}
	\int^{\perp}_{s,t,s',t'}\frac{\Theta(X_0-|s_{\perp}|)\Theta(Y_0-|t_{\perp}|)\Theta(|s_{\perp}|-|s'_{\perp}|)\Theta(|t_{\perp}|-|t'_{\perp}|)}{|s_{\perp}||t_{\perp}||s'_{\perp}||t'_{\perp}|}
	\\\nonumber
	&&\times\big(|u'_{\perp}|-|u_{\perp}|\big)\big(|t'_{\perp}|-|t_{\perp}|\big)
	\big[\hat{s}^y_{\perp}\hat{t}^y_{\perp}\partial^2_{u'x}\partial^2_{v'x}-2\hat{s}^y_{\perp}\hat{t}^x_{\perp}\partial^2_{u'x}\partial_{v'y}\partial_{v'x}
	\\
	&&+\hat{s}^x_{\perp}\hat{t}^x_{\perp}\partial_{u'x}\partial_{u'y}\partial_{v'x}\partial_{v'y}\big]
	\Omega_{-}(u_{\perp},v_{\perp})\Omega_{+}(u'_{\perp},v'_{\perp}).
\end{eqnarray}

\section{Semi-analytical analysis of the integrals ${\cal I}_{1},{\cal I}_{2}, {\cal I}_{3}$}
\label{app:alternative_I}
We now analyse the behavior of different integrals. We first introduce $\rho=Q_s^2|u_{\perp}-v_{\perp}|^2$ and accordingly write GBW distribution
\begin{eqnarray}
\Omega_{\pm}(u_{\perp},v_{\perp})=\Omega(\rho)=\frac{Q_s^4}{g^4N_c^2}\left(\frac{1-e^{-\rho/4}}{\rho/4}\right)^2, \label{eq:omegarho}
\end{eqnarray}
similarly $\Omega_{\pm}(u'_{\perp},v'_{\perp})=\Omega(\rho')$ with $\rho'=Q_s^2|u'_{\perp}-v'_{\perp}|^2$. Using Eq. (\ref{eq:omegarho}) we can obtain
\begin{eqnarray}
\partial_{ui}\Omega_{\pm}(\rho)=(\partial_{ui}\rho)\Omega'_{\pm}(\rho')=2Q_sw^i\Omega'_{\pm}(\rho)=-\partial_{vi}\Omega_{\pm}(\rho),
\end{eqnarray}
\begin{eqnarray}
\partial_{vj}\partial_{ui}\Omega_{\pm}(\rho)=-2Q_s^2\big(\delta^{ji}\Omega_{\pm}'(\rho)+2w^iw^j\Omega_{\pm}''(\rho)\big),
\end{eqnarray}
\begin{eqnarray}\nonumber
\partial_{uk}\partial_{vj}\partial_{ui}\Omega_{\pm}(\rho)&=&-\partial_{vk}\partial_{vj}\partial_{ui}\Omega_{\pm}(\rho)
\\
&=&-4Q_s^3\big[\big(\delta^{ji}w^k+\delta^{kj}w^i+\delta^{ki}w^j\big)\Omega_{\pm}''(\rho)
+2w^iw^jw^k\Omega_{\pm}'''(\rho)\big],
\end{eqnarray}
and
\begin{eqnarray}\nonumber
\partial_{vl}\partial_{uk}\partial_{vj}\partial_{ui}\Omega_{\pm}(\rho)&=&\partial_{ul}\partial_{uk}\partial_{uj}\partial_{ui}\Omega_{\pm}(\rho)
\\\nonumber
&=&4Q_s^4\big[\big(\delta^{ji}\delta^{kl}+\delta^{kj}\delta^{il}+\delta^{ki}\delta^{jl}\big)\Omega_{\pm}''(\rho)+2\big(w^i(w^j\delta^{kl}+w^k\delta^{jl}+w^l\delta^{jk})
\\
&&+w^k(w^l\delta^{ij}+w^j\delta^{il})+\delta^{ik}w^jw^l\big)\Omega_{\pm}'''(\rho)
+4w^iw^jw^kw^l\Omega_{\pm}''''(\rho)\big],
\end{eqnarray}
where $w^i=Q_s(u_{\perp}^i-v_{\perp}^i)$. 

\begin{figure}
	%\vspace{-1cm}
	\begin{center}
		\includegraphics[width=0.5\hsize]{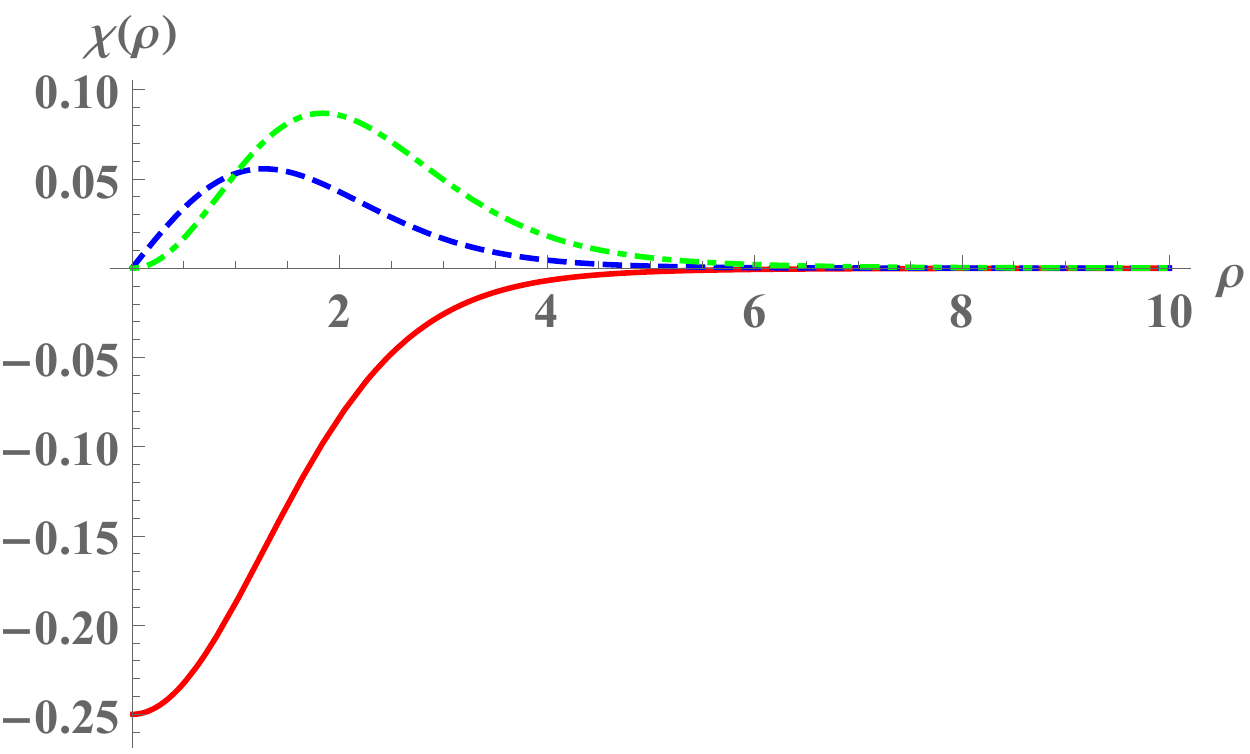}
	\end{center}
	%\vspace{-1cm}
	\caption{The solid red, dashed blue, and dashed-dotted green curves correspond to $\Omega'(\rho)$, $\rho^{1/2}\Omega''(\rho)$, and $\rho\Omega''(\rho)$ normalized by $\frac{Q_s^4}{g^4N_c^2}$, respectively.
	}
	\label{DOmega_plot}
\end{figure}

The above equations can be utilized to obtain
\begin{eqnarray}
\partial_{vx}\partial_{ux}\Omega_{\pm}(\rho)=-2Q_s^2\big(\Omega_{\pm}'(\rho)+2(w^x)^2\Omega_{\pm}''(\rho)\big),
\end{eqnarray}
\begin{eqnarray}
\partial_{vy}\partial^2_{ux}\Omega_{\pm}(\rho)=-4Q_s^3\big[\omega^y\Omega_{\pm}''(\rho)
+2(w^x)^2w^y\Omega_{\pm}'''(\rho)\big],
\end{eqnarray}
\begin{eqnarray}
\partial_{vx}\partial^2_{ux}\Omega_{\pm}(\rho)=-\partial_{ux}\partial^2_{vx}\Omega_{\pm}(\rho)=-4Q_s^3\big[3w^x\Omega_{\pm}''(\rho)
	+2(w^x)^3\Omega_{\pm}'''(\rho)\big],
\end{eqnarray}
\begin{eqnarray}
\partial_{vy}\partial_{vx}\partial_{ux}\Omega_{\pm}(\rho)
	&=&4Q_s^3\big[w^y\Omega_{\pm}''(\rho)
	+2w^y(w^x)^2\Omega_{\pm}'''(\rho)\big],
\end{eqnarray}
\begin{eqnarray}
	\partial_{vy}(\partial_{ux})^3\Omega_{\pm}(\rho)=-\partial_{vy}\partial_{vx}(\partial_{ux})^2\Omega_{\pm}(\rho)=4Q_s^4\big[6w^xw^y\Omega_{\pm}'''(\rho)
	+4(w^x)^3w^y\Omega_{\pm}''''(\rho)\big],
\end{eqnarray}
\begin{eqnarray}
\partial_{vx}\partial_{ux}\partial_{vy}\partial_{uy}\Omega_{\pm}(\rho)=4Q_s^4\big[\Omega_{\pm}''(\rho)+2\big((w^x)^2+(w^y)^2\big)\Omega_{\pm}'''(\rho)
+4(w^y)^2(w^x)^2\Omega_{\pm}''''(\rho)\big],
\end{eqnarray}
and
\begin{eqnarray}
	(\partial_{vx})^2(\partial_{ux})^2\Omega_{\pm}(\rho)
	&=&4Q_s^4\big[3\Omega_{\pm}''(\rho)+12(w^x)^2\Omega_{\pm}'''(\rho)
	+4(w^x)^4\Omega_{\pm}''''(\rho)\big].
\end{eqnarray}
Finally, one can find
\begin{eqnarray}\nonumber
	\mathcal{I}_{1}&\approx& -2Q_s^2\frac{g^4N_c^2(N_c^2-1)}{4(2\pi)^4p_0^4}
	\int^{\perp}_{s,t,s',t'}\frac{\Theta(X_0-|s_{\perp}|)\Theta(Y_0-|t_{\perp}|)\Theta(|s_{\perp}|-|s'_{\perp}|)\Theta(|t_{\perp}|-|t'_{\perp}|)}{|s_{\perp}||t_{\perp}||s'_{\perp}||t'_{\perp}|}
	\\
	&&\times \big[\hat{s}_{\perp}^{x}\hat{t}_{\perp}^{x}\hat{s}_{\perp}^{\prime y}\hat{t}_{\perp}^{\prime y}+\hat{s}_{\perp}^{y}\hat{t}_{\perp}^{y}\hat{s}_{\perp}^{\prime x}\hat{t}_{\perp}^{\prime x}-2\hat{s}_{\perp}^{x}\hat{t}_{\perp}^{y}\hat{s}_{\perp}^{\prime y}\hat{t}_{\perp}^{\prime x}\big]\big(\Omega'(\rho)+2(w^x)^2\Omega''(\rho)\big)
	\Omega(\rho'),
\end{eqnarray}

\begin{eqnarray}\nonumber
	\mathcal{I}_{2}
	&\approx& 4Q_s^3\frac{g^4N_c^2(N_c^2-1)}{2(2\pi)^4p_0^4}
	\int^{\perp}_{s,t,s',t'}\frac{\Theta(X_0-|s_{\perp}|)\Theta(Y_0-|t_{\perp}|)\Theta(|s_{\perp}|-|s'_{\perp}|)\Theta(|t'_{\perp}|-|t_{\perp}|)}{|s_{\perp}||t_{\perp}||s'_{\perp}||t'_{\perp}|}
	\\\nonumber
	&&\times\big(|t_{\perp}|-|t'_{\perp}|\big)\hat{s}_{\perp}^{[x} \hat{s}'^{ y]}_{\perp}\big[\hat{t}'^y_{\perp}\big(3w^x\Omega''(\rho)
	+2(w^x)^3\Omega'''(\rho)\big)-\hat{t}'^x_{\perp}\big(w^y\Omega''(\rho)
	+2w^y(w^x)^2\Omega'''(\rho)\big)\big]\Omega(\rho').
	\\
\end{eqnarray}
and
\begin{eqnarray}\nonumber
	\mathcal{I}_{3}
	&\approx& -4Q_s^4\frac{g^4N_c^2(N_c^2-1)}{4(2\pi)^4p_0^4}
	\int^{\perp}_{s,t,s',t'}\frac{\Theta(X_0-|s_{\perp}|)\Theta(Y_0-|t_{\perp}|)\Theta(|s_{\perp}|-|s'_{\perp}|)\Theta(|t_{\perp}|-|t'_{\perp}|)}{|s_{\perp}||t_{\perp}||s'_{\perp}||t'_{\perp}|}
	\\\nonumber
	&&\times\big(|u'_{\perp}|-|u_{\perp}|\big)\big(|t'_{\perp}|-|t_{\perp}|\big)
	\big[\hat{s}^y_{\perp}\hat{t}^y_{\perp}\big(3\Omega''(\rho')+12(w'^x)^2\Omega'''(\rho')
	+4(w'^x)^4\Omega''''(\rho')\big)
	\\\nonumber
	&&+2\hat{s}^y_{\perp}\hat{t}^x_{\perp}\big(6w'^xw'^y\Omega'''(\rho')
	+4(w'^x)^3w'^y\Omega''''(\rho')\big)
	\\
	&&+\hat{s}^x_{\perp}\hat{t}^x_{\perp}\big(\Omega''(\rho')+2\big((w'^x)^2+(w'^y)^2\big)\Omega'''(\rho')
	+4(w'^y)^2(w'^x)^2\Omega''''(\rho')\big)\big]
	\Omega(\rho).
\end{eqnarray}
where $w'^i=Q_s(u'^i_{\perp}-v'^{i}_{\perp})$. Recall that $(u_{\perp}^i-v_{\perp}^i)=(t_{\perp}^i-s_{\perp}^i)$ and $(u'^i_{\perp}-v'^i_{\perp})=(t'^i_{\perp}-s'^i_{\perp})$ when $r_{\perp}=0$. Here $\Omega'(\rho)\equiv \partial_{\rho}\Omega(\rho)$ and similar notations are applied to higher derivatives. Noticeably, a derivative acting on $\Omega$ flips the overall sign even though the magnitudes of $\Omega(\rho)$ and its higher derivatives monotonically decrease with $\rho$. However, the prefactors with different orders of $w^i$ could qualitatively convert the monotonically decreasing (or increasing) function into an approximate pulse form with a shift of the maximum to larger $\rho$. One can see this behavior in Fig.~\ref{DOmega_plot}. Consequently, despite the angular dependence, when having larger $X_0$ and hence larger phase space for the integrals, $\mathcal{I}_{1,2,3}$ may not monotonically increase due to the combination of terms with different derivatives and accompanied prefactors with distinct orders of $\omega^i$ (or $\omega'^i$). As illustrated in Fig.~\ref{figI123}, we have observed the turnover of $\hat{\mathcal{I}}_{2}$, while the non-monotonic behaviors of $\hat{\mathcal{I}}_{1,3}$ occurs at much larger $Q_sX_0$.
%\newpage
%\bibliographystyle{apsrev4-1}
\bibliography{spin_correlation_CGC_PRDfinal.bbl}
%\bibliography{polarization_ref}
%\input{polarization_ref}  
\end{document}